# Investment Portfolio Optimization Based on Modern Portfolio Theory and Deep Learning Models

Maciej Wysocki[1], Paweł Sakowski[2]

[1] Quantitative Finance Research Group, Department of Quantitative Finance and Machine Learning, Faculty of Economic Sciences, University of Warsaw, University of Warsaw, ul. Długa 44/50, 00-241 Warsaw, Poland. Corresponding author: m.wysocki9@uw.edu.pl

[2] Quantitative Finance Research Group, Department of Quantitative Finance and Machine Learning, Faculty of Economic Sciences, University of Warsaw, University of Warsaw, ul. Długa 44/50, 00-241 Warsaw, Poland. E-mail: p.sakowski@uw.edu.pl

**Abstract:** This paper investigates an important problem of an appropriate variance-covariance matrix estimation in the Modern Portfolio Theory. We propose a novel framework for variance-covariance matrix estimation for purposes of the portfolio optimization, which is based on deep learning models. We employ the long short-term memory (LSTM) recurrent neural networks (RNN) along with two probabilistic deep learning models: DeepVAR and GPVAR to the task of one-day ahead multivariate forecasting. We then use these forecasts to optimize portfolios of stocks and cryptocurrencies. Our analysis presents results across different combinations of observation windows and rebalancing periods to compare performances of classical and deep learning variance-covariance estimation methods. The conclusions of the study are that although the strategies (portfolios) performance differed significantly between different combinations of parameters, generally the best results in terms of the information ratio and annualized returns are obtained using the LSTM-RNN models. Moreover, longer observation windows translate into better performance of the deep learning models indicating that these methods require longer windows to be able to efficiently capture the long-term dependencies of the variance-covariance matrix structure. Strategies with less frequent rebalancing typically perform better than these with the shortest rebalancing windows across all considered methods.

**Note:** This research did not receive any specific grant from funding agencies in the public, commercial, or not-for-profit sectors.

# 1 Introduction

Portfolio management and optimal funds allocation remains a subject of interest for any person or institution actively trading financial assets. The growing access to technology and computational resources combined with development in technology and science brought a new era to many fields, including finance, hence the tools used to make investment decisions changed significantly. Investors nowadays may easily use techniques ranging from simple analyses based on financial data to sophisticated portfolio selection frameworks based on deep reinforcement learning to arrive at an allocation that fulfills their requirements (Lucarelli and Borrotti, 2020; Zhang, Zohren and Roberts, 2020).

The concept of portfolio involves holding any positions in at least two different assets, without a limitation regarding the type of position, asset, quantity or even market. Portfolio optimization is therefore a procedure of constructing a portfolio with the highest possible returns for a desired level of risk given the universe of available assets. It was first introduced by Harry Markowitz (1952) and goes by the name of Modern Portfolio Theory (MPT). Its main point leverages the idea of diversification, which is generally superior over holding a position in a single asset since it enables achieving higher returns for the same level of risk (Melicher and Norton, 2013).

Although MPT seems temptingly simple at first glance, its practical implementation requires several assumptions and additional steps (Zhang, Li and Guo, 2018), the most complicated of which is estimation of the assets' variance-covariance matrix. To address this issue, numerous solutions have already been proposed, including but not limited to: traditional frequentist estimator (Georgiev, 2014; Xidonas et al., 2020), exponentially weighted moving average (EWMA) (Xidonas et al., 2020; Henriques and Ortega, 2014), BEKK-GARCH by Baba et al. (1990) (Naccarato and Pierini, 2014; Zakamulin, 2015), dynamic conditional correlation (DCC) by Engle (2002) (Moura, Santos and Ruiz, 2020; Henriques and Ortega, 2014; Zakamulin, 2015), linear and non-linear shrinkage estimators (Lam, 2020; Frahm and Memmel, 2010) and support vector regression (Fiszeder and Orzeszko, 2021). Another approach to variance-covariance matrix was presented in Albuquerque et al. (2023), where authors tested their method with a variety of portfolio optimization techniques.

The recent studies show that applications of the deep learning time series forecasting methods to financial data might provide advantage over the traditional econometric methods in fields such as forecasting stock prices and indices, forex trading, modelling volatility etc. (see e.g., Sezer et al., 2020 or Huang et al., 2020 for an extensive literature review). As of 2022, the most popular models used for financial time series are recurrent neural networks, in particular Long Short-Term Memory (LSTM) neural networks, and transformers with attention mechanisms (see e.g. Ding et al., 2020 or Makridakis et al., 2020). The up-to-date research provides evidence for improved forecasting accuracy of global, probabilistic models compared with the local approach, especially for multivariate time series (Salinas et al., 2019; Salinas et al., 2020).

In this study we present an application of state-of-the-art deep learning and probabilistic deep learning models to forecasting variance-covariance matrices, which are used for portfolio optimization in the MPT framework. The probabilistic global models such as Gaussian Process VAR (GPVAR) or DeepVAR used in this study allow to model a whole distribution hence providing a broader insight into the phenomenon (Alexandrov et al., 2020). We compare the performances of such portfolios with the others based on the classical approaches such as frequentist estimator or EWMA.

In order to provide comprehensive evidence of the performance of our framework, we use a universe of assets consisting of stocks and cryptocurrencies to form the portfolios. The novel asset class is capable of enhancing the diversification of portfolios (Demiralay and Bayracı, 2020) as well as providing superior financial results (Li et al., 2021) compared to the portfolios based only on stocks. In general, cryptocurrencies are considered to be more volatile assets than stocks and other classical assets typically used for constructing portfolios (Ghorbel and Jeribi, 2021), hence including them provides an additional opportunity to test the forecasting performance of deep learning based models in varying market conditions.

The goal of this paper is therefore to introduce a framework for portfolio optimization combining the MPT with deep learning models and present an extensive comparison of investment strategies utilizing that framework and probabilistic DL models. Both of those shall enrich the literature by providing a ground for further research in this area as well as serve the market practitioners. The research hypotheses of this study are as follows:

1. The strategies utilizing the variance-covariance matrix estimations from the deep learning methods outperform the strategies based on the classical variance-covariance matrix estimation methods.

2. The strategies based on the variance-covariance matrix estimations from the probabilistic deep learning (i.e. GPVAR od DeepVAR). models perform better than the strategies based on the simple LSTM-RNN models.

The remaining parts of the paper are structured as follows. Section 2 contains an explanation of methodology and models used in the study. Section 3 covers the overview of portfolios construction and evaluation methodology. Section 4 presents the description of data used in the study. The empirical results are presented in Section 5. The last part of the paper contains conclusions and further research ideas. The codes used to arrive at results presented in this paper were developed using Python 3.8 and the deep learning frameworks used for neural networks development were Tensorflow 2.2 and GluonTS. A lot of effort has been into reproducibility, hence these are available at the GitHub repository: github.com/Maciej-13/vcov_forecast.

## 2 Theoretical Background

### 2.1 Mean-Variance Optimization

The fundamentals of the Modern Portfolio Theory were laid down by Markowitz (1952), when he created a Mean-Variance Optimization framework for portfolio optimization. A rational economic agent maximizes utility by choosing a portfolio with the highest returns for a given level of risk or, likewise, the lowest risk for a given level of return. In the mean-variance model, risk and returns were reflected respectively by the portfolio's variance and expected returns, which are typically proxied by mean historical returns. The portfolio allocation problem can be expressed as the following optimization task:

$$min \sum_{i=1}^{n} \sum_{j=1}^{n} \omega_i R_i \omega_j R_j \sigma_{ij}$$

$$max \sum_{i=1}^{n} \omega_i \mu_i$$

$$s.t. \begin{cases} \sum_{i=1}^{n} \omega_i = 1 \\ \forall_{i \epsilon \{1,...,n\}} \omega_i \geq 0 \end{cases}$$

where $R$ is a vector of multivariate returns, $\omega_i$ is share of asset $i$ in the portfolio, $\mu$ is a vector of expected returns and $\sigma_{ij}$ is a covariance between assets $i$ and $j$. This version of the mean-variance model does not allow shorting stocks, meaning that exclusively long-only portfolios are considered in this study.

As noted earlier, although the framework is simple and easily understandable, there are multiple factors to be addressed when utilizing the MPT for real-world investment applications. First of all, the choice of optimization objective has to be made. In this study, minimization of variance is selected and used for all, which is rather a stable choice for portfolios incorporating cryptocurrencies (Brauneis and Mestel, 2019). There are however numerous other possibilities such as e.g., Sharpe ratio optimization (Sharpe, 1998). Sakowski et al. (2016) provided an overview of optimization criteria for Markowitz model in algorithmic investment strategies. Secondly, a method of expected returns calculations must be chosen and, in this case, mean historical returns are used. The last decision concerns the risk model, namely the risk quantification for the mean-variance optimization framework, which is the topic of this study.

## 2.2 Classical Variance-Covariance Estimators

In this paper, we refer to all non deep learning-based variance-covariance matrix forecasting methods as the classical estimators, although some of them are newer than the others.

The first approach is the frequentist estimator of the variance-covariance matrix, namely the sample variance-covariance matrix. In the resulting matrix, the entrances are sample covariances of all possible asset pairs and assets' variances on the diagonal:

$$\hat{\Sigma}_t = \frac{1}{k-1} \sum_{m=0}^{k} (R_i^{t-m} - \bar{R}_i)(R_j^{t-m} - \bar{R}_j)$$

where $R_i^{t-m}$ denotes the returns of asset $i$ in the period $[t-k, t]$, $\bar{R}_i$ is the average of returns and $k$ is the window parameter controlling how many past observations are considered in the calculations. Although this estimator possesses some attractive theoretical properties since it is an unbiased estimator of the variance-covariance matrix as well as a maximum likelihood estimator under the assumption of normality (Ledoit and Wolf, 2003). Although it is also very easy to obtain from any set of assets, it comes with very little structure and causes multiple problems when used in practice (Jobson and Korkie, 1980). If the number of stocks is larger than the number of time steps in the data, then the estimated matrix will be singular, while the true variance-covariance matrix should not be singular.

The second approach is only a slight modification of the sample variance-covariance estimator, namely the semi-covariance matrix, which is also based on calculation of sample covariances and variances, but only returns below certain threshold are taken into calculations (Estrada, 2008):

$$\hat{\Sigma} = \frac{1}{k} \sum_{m=0}^{k} Min(R_i^{t-m} - B, 0) * Min(R_j^{t-m} - B, 0)$$

where parameter $B$ is the returns threshold. An additional choice of this benchmark for returns is required and, in this paper, we use a 2% threshold of returns for the semi-variance estimator. This method is focused more on the downside risk, which concerns investors more than any over-performance and Markowitz himself referred to semi-variance as superior to variance in portfolio creations (Markowitz, 1991).

Another variance-covariance estimator is the exponentially weighted variance-covariance matrix, which is based on the idea of assigning more weights to the recent data, since intuitively the newer observations are more relevant to the present state. The exponentially weighted variance-covariance matrix estimator is given with the formula:

$$\hat{\Sigma}_t = \lambda \hat{\Sigma}_{t-1} + (1 - \lambda)(R_t - \mu)(R_t - \mu)'$$

where $\lambda$ is a decay rate and $\mu$ is a vector of the expected returns. Following the RiskMetrics (1996) EWMA methodology for daily returns, the decay factor of 0.94 is chosen for this estimator in this study.

The next class of the variance-covariance matrix estimators used in this study are shrinkage estimators, which are mainly based on Ledoit and Wolf methodologies (Ledoit and Wolf, 2001, 2003). The general idea is to compose an estimator as a combination of a highly structured estimator (i.e. with a small number of parameters) $F$ called the shrinkage target and an unstructured sample variance-covariance estimator $S$. Such composition is given as:

$$\hat{\Sigma} = \delta F + (1 - \delta)S, \quad 0 \leq \delta \leq 1$$

where $\delta$ is a shrinkage coefficient. This combination benefits from a low bias of the unstructured estimator and a low error of the structured one. There are however multiple possibilities of structured estimators to choose from and each of them is characterized by a different optimal shrinkage coefficient. In this study the following three structured estimators were used: constant variance shrinkage, single factor shrinkage (Ledoit and Wolf, 2001) and constant correlation shrinkage (Ledoit and Wolf, 2003). The constant variance shrinkage uses a diagonal target matrix with covariances equal to zero and variances equal to mean asset's variance. The single factor shrinkage uses a single-index variance-covariance matrix estimator, which is based on the Sharpe's single index model (Sharpe, 1963). The last method, called the constant correlation shrinkage aims to set all asset pairs' correlations to the mean of all sample correlations, while the sample variances are not modified.

The last classical estimator in this study is the oracle approximating estimator (Chen et al., 2010). It is also based on the shrinkage approach, but it is supposed to provide a lower mean squared error and therefore a better variance-covariance estimation than other shrinkage estimators if the sample is Gaussian or near-Gaussian. This estimator was designed specifically to address some weaknesses of Ledoit and Wolf shrinkage estimators in particular cases with a high number of dimensions and a small sample. This method uses an iterative approach to compute a modified shrinkage coefficient that is superior to other shrinkage estimators under the assumption of sample normality. Calculation of the shrinkage estimator can be expressed as:

$$\hat{\rho}_{t+1} = \frac{\left(1 - \frac{2}{p}\right) * Tr(\hat{\Sigma}_t S) + Tr^2(\hat{\Sigma}_t)}{\left(n + 1 - \frac{2}{p}\right) * Tr(\hat{\Sigma}_t S) + \left(1 - \frac{n}{p}\right) * Tr^2(\hat{\Sigma}_t)}$$

$$\hat{\Sigma}_{t+1} = (1 - \rho_{t+1})S + \rho_{t+1}F$$

$$F = \frac{Tr(S)}{p}\mathbf{I}$$

where $S$ is the sample variance-covariance matrix, $p$ stands for the number of dimensions and $Tr(\cdot)$ is a trace of the matrix.

## 2.3 Long Short-Term Memory Neural Network

Long Short-Term Memory (LSTM) neural network is an architecture of recurrent neural networks (RNN) designed specifically to process sequential data such as time series or sequences of words (Hochreiter and Schmidhuber, 1997). The classical RNNs with a cyclical architecture operating on ordered arrays of inputs face the vanishing gradient problem (Hochreiter, 1998), which eventually leads to inability of learning the long-term dependencies (Kolen and Kremer, 2001). Therefore, the LSTM design relies on use of cells composed of gates that allow to incorporate both current and previous information during a single data processing and tackle the optimization-related problem.

Figure 1. Architecture of a LSTM unit.

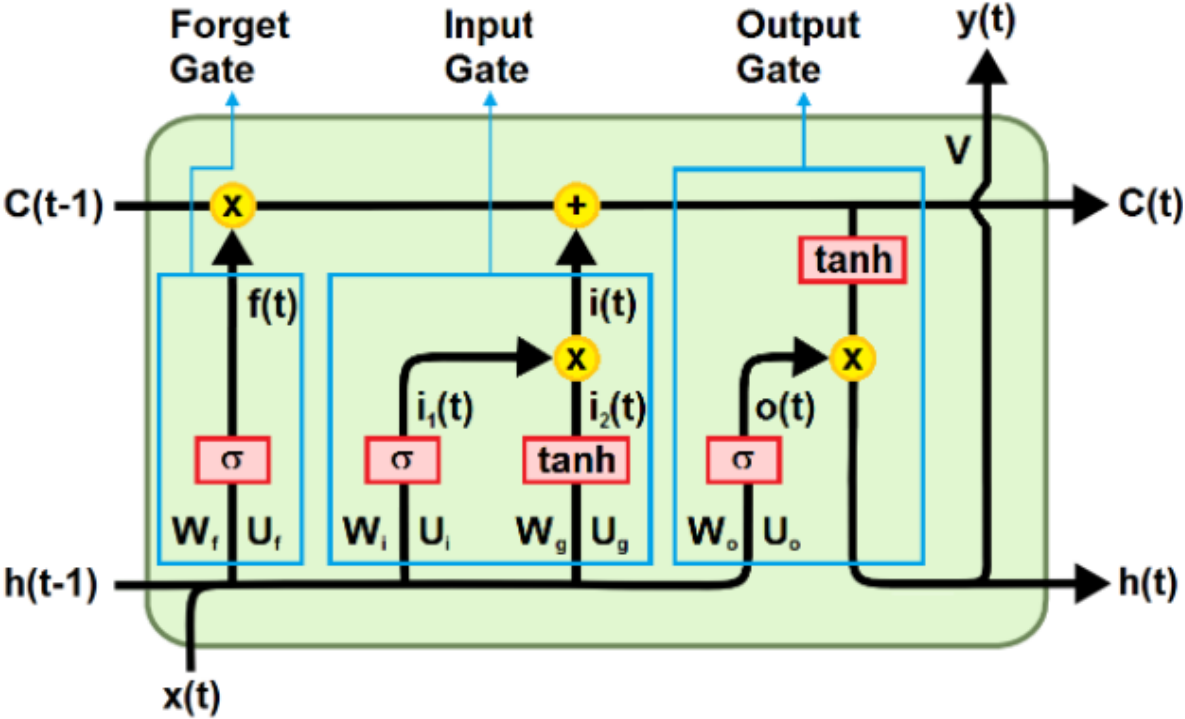

*Source:* Image by Marco Del Pra downloaded from: towardsdatascience.com/time-series-forecasting-with-deep-learning-and-attention-mechanism-2d001fc871fc. Accessed on 01-05-2022.

Figure 1 presents the architecture of a standard LSTM cell, and it clearly exhibits four basic components of a unit: cell state, forget gate, input gate and output gate. The cell state is the actual sequence representing the memory of the RNN. The first part inside the unit is the forget gate, which is responsible for deciding which past information should be preserved or removed, hence its name related to forgetting. The input gate oversees processing the current data, that is extracting and retaining the relevant information. The final, output gate calculates the current timestamp's output and passes it outside the cell. The general connections are defined as follows: the input vector $x_t$ is connected with the unit by matrix $U$, the unit is connected to the previous $(t-1)$ and next $(t+1)$ unit with a matrix $W$ and finally the output vector $y_t$ is connected with the unit by a matrix $V$. Matrices $U, W$ and $V$ store weights which are optimized during the learning process and are distinguished for each of the unit parts. The LSTM cell architecture can therefore be described as the following set of equations:

$$f_t = \sigma(W_f h_{t-1} + U_f x_t + b_f)$$
$$i_{1,t} = \sigma(W_i h_{t-1} + U_i x_t + b_i)$$
$$i_{2,t} = \tanh(W_g h_{t-1} + U_g x_t + b_g)$$
$$i_t = i_{1,t} \circ i_{2,t}$$
$$c_t = \sigma(f_t c_{t-1} + i_t)$$
$$o_t = \sigma(x_t U_o + h_{t-1} W_o + b_o)$$
$$h_t = \tanh(c_t) \circ o(t)$$
$$\hat{y}_t = V h_t$$

where $W_f, W_i, W_g, W_o, U_f, U_i, U_g, U_o, V$ are appropriate weight matrices of adequately forget gate $(f)$, input gate $(i)$, cell state $(g)$, output gate $(o)$ and output vector, $b_f, b_i, b_g, b_o$ are biases of each gate, $h_t$ is the hidden state, $x_t$ is the input to the LSTM unit, $c_t$ is the cell state, $\sigma$ is the sigmoid activation function and the symbol ∘ denotes the Hadamard product.

## 2.4 Probabilistic Autoregressive Recurrent Neural Networks

Probabilistic forecasting is a field of time series analysis focused on modelling a probabilistic distribution of data and therefore such methods are referred to as global models, since they try capture the holistic view of the considered phenomenon. Moreover, the probabilistic time series forecasting models for multivariate tasks considered in this study estimate a combined distribution of all series included in the dataset, rather than learning and predicting their nature separately. Therefore, such models are better suited for tasks when interdependencies between the time series might play a significant role as well as for forecasting exercises with many long series, hence modelling and forecasting variance-covariance matrices for portfolio optimization is an appropriate problem for application of the probabilistic deep learning models. Two LSTM-based probabilistic neural networks architectures are used in this study, namely DeepVAR (Salinas et al., 2020) and Gaussian Process VAR (GPVAR) (Salinas et al., 2019).

DeepVAR is a multivariate probabilistic deep learning model for time series based on the LSTM cells, which main idea is estimating the conditional distribution of time series given their preceding values:

$$P(z_{t:T}|z_{1:t-1}, x_{1:T})$$

where $z$ is a matrix of time series, $[1, t-1]$ and $[t, T]$ are respectively conditioning and prediction ranges, and $x$ is a matrix of covariates. Both ranges' definitions could potentially differ between the forecasted series, but in the case of modelling variance-covariance matrices we used the same value of $t$ for every entrance. The model distribution is expressed as a product of likelihood functions $l(z|\theta)$, where $\theta$ is vector of parameters, which depends on the outputs of the autoregressive RNN. DeepVAR is both autoregressive and recurrent, as during the training process each time stamp is estimated using lagged observations and the previous output of the NN as the inputs. The output from RNN is used to calculate the parameters of the likelihood function and update the model parameters. In this study, we chose Gaussian likelihood as the covariances could be both positive and negative:

$$l(z|\mu, \sigma) = \frac{1}{\sqrt{2\pi}\sigma} e^{-\frac{(z-\mu)}{2\sigma^2}}$$

where $\mu$ and $\sigma$ are parameters of the multivariate normal distribution and are both defined as functions of the RNN output.

GPVAR is another multivariate probabilistic deep learning time series model estimating the joint conditional distribution within an LSTM-based autoregressive RNN. The joint distribution is parametrized using a Gaussian copula process, which parameters depend on the model state:

$$h_{i,t} = \varphi_{\theta_h}(h_{i,t-1}, z_{i,t-1})$$

$$P(z_t|h_t) = \left( [f_1(z_{1,t}), f_2(z_{2,t}), \dots, f_N(z_{N,t})]^T \middle| \mu(h_t), \Sigma(h_t) \right)$$

where $h_t$ is state of the model with transition dynamic $\varphi$, $\mu$ and $\Sigma$ are parameters of the Gaussian distribution and $f_i$ are functions of form: $\Phi^{-1} \circ \hat{F}_i$ combining the inverse of the standard normal distribution cumulative distribution function (CDF) and empirical marginal distribution of $i$-th input series. These functions' purpose is transformation of each time series to marginally follow a standard Gaussian distribution. Under such setting, the CDF of the joint distribution of the input time series is a Gaussian copula. All parameters of transition dynamics, $\mu$ and $\Sigma$ are estimated from the data during the learning process which is conducted by minimization of the likelihood function. Estimation of the distribution parameters could be done with an additional parametrization of the distribution variance-covariance matrix, as it is expressed as a sum of diagonal and low-rank matrices to make the calculations more efficient. All model parameters are updated jointly during the training process, therefore authors of the model expressed it as a parametrization of a Gaussian process to allow learning on a subset of inputs, rather than using all series, which makes the training less computationally expensive.

Both methods provide forecasts in the form of Monte Carlo samples, therefore enable a deeper analysis of predicted values as well construction of confidence intervals around the forecasts. In this study we decided to use median values as the forecasts, as it is a middle quantile and a commonly undertaken approach in probabilistic forecasting (Tyralis and Papacharalampous, 2022).

## 2.5 Forecasting Variance-Covariance Matrix with Deep Learning Methods

Forecasting the variance-covariance matrix with any non-traditional method requires some preparation to ensure that the outputs are interpretable and reflect the expected variance-covariance matrix structure. In general, any variance-covariance matrix is symmetric and positive-semidefinite, which can be expressed as:

$$\Sigma = \Sigma^T$$

$$\forall_{x \epsilon \mathbb{R}^n} x^T \Sigma x \geq 0$$

A variance-covariance matrix created as a combination of forecasts of its entrances is not guaranteed to meet these conditions and therefore it would not be possible to use such matrix for the portfolio optimization process. To impose a proper structure on the forecasted matrix, a Cholesky decomposition is commonly used, especially in financial applications (Chiriac, et al., 2011; Bucci, 2020; Fiszeder and Orzeszko, 2021). This factorization method decomposes a positive-definite matrix into a product two matrices, one of which is lower (or upper) triangular and the second is a conjugate transposition of the first one:

$$A = XX^*$$

where $A$ is a Hermitian, positive-definite matrix, $X$ is a lower triangular matrix, which on the diagonal has real, positive values and $X^*$ is its conjugate transposition. In the case of positive-semidefinite matrices, as the variance-covariance matrices, this decomposition still applies, but might not be unique. The Cholesky factor can be transformed back into the original matrix by a simple matrix multiplication, therefore allowing to reconstruct the matrix preserving its initial properties.

The variance-covariance matrix forecasting methodology for deep learning models used in this paper can be summarized as a list of the following steps:

1. For each available timestamp calculate the historical $N \times N$ variance-covariance matrix $\Sigma_t$ over the selected window $w$ (where $N$ is the number of assets).
2. Apply the Cholesky decomposition to each of the obtained variance-covariance matrices $\Sigma_t = X_t X_t^*$
3. Construct separate time series $x_t^{i,j}$ for each of the entrances of the decomposed matrices resulting in $\frac{N(N+1)}{2}$ series.
4. Forecast the obtained series of Cholesky factors' entrances using a selected deep learning method trained on the available observations.
5. Construct the Cholesky factors $X_{T+n}$ from the forecasted series and then reconstruct the variance-covariance matrix $\Sigma_{T+n} = X_{T+n} X_{T+n}^*$.

The assumed time range is $[1, 2, \ldots, t, \ldots, T]$, where $T$ is the last observed timestamp and $T + n$ is the forecast horizon. The number of assets is $N$ and in total it is required to model $\frac{N(N+1)}{2}$ series, which are variances and covariances of assets. For the purposes of the investment strategies based on the portfolio optimization only a single variance-covariance matrix estimation is required, therefore we used a one-step ahead forecast ($n = 1$). Therefore, the in-sample periods are determined by the arbitrarily chosen window length (30, 60, 90, 120), while the out-of-sample period is always one day and this forecast is an input to the portfolio optimization process.

**3 Investment Strategies**

## 3.1 Portfolio Construction

The portfolios in this study were created within the Markowitz framework using various variance-covariance estimation methods and different combinations of parameters. However, the construction methodology was shared across these strategies to ensure that the results are comparable. All portfolios were optimized using the minimum variance criterion, that is the optimization goal was to search for weights which provide the most stable portfolio in terms of the provided variance-covariance matrix. An additional objective function was used for portfolios with weights rebalancing that incorporated a simple model of transactional costs to ensure that the cost of changing positions is taken into consideration during weights optimization. For small portfolios the transactional costs may be negligible, but the larger the portfolio, the higher are the costs, therefore in this study we have decided to use such solution and consider all costs of changing position assuming a fixed percentage commission of 50 bps for every asset. Although this solution is a simplification, it allows to reflect the transactional costs in this research setup and simulate the real-world trading circumstances.

The single portfolio optimization process was performed as follows:

1. Gather the available prices of assets and calculate the expected returns estimated as the mean historical returns for a given historical window.
2. Estimate the variance-covariance matrix using the selected method and optimize weights for a given rebalancing period.
3. Calculate discrete portfolio allocation from the optimal weights, allocate the available capital to update the portfolio structure by buying and selling appropriate assets.

The portfolio optimization steps were done on every rebalancing day (arbitrarily chosen, every 30, 60, 90 and 120 days) and between these dates only the current portfolio value was calculated and saved. A greedy discrete allocation algorithm, that operates in two stages, was used to transform weights into market orders. First as many shares or coins as possible within weights were bought for the available cash and each time the number of assets was rounded down to the closest integer. Then, for each asset the difference between the current share in portfolio and the desired optimal weight was calculated and the assets with the highest difference were bought first and the weights deviations calculation was conducted again. This process continued until no more available cash could be invested, leaving very little cash on hand, and allowing to get very close to the optimal portfolio allocation.

## 3.2 Performance Measures

To compare and evaluate the investment strategies we employed multiple performance measures to get a broad overview of strategies behavior:

- annualized rate of returns:

$$aRC = \left(1 + \frac{P_T}{P_0}\right)^{\frac{365}{T}} - 1$$

- annualized standard deviation of returns:

$$aSD = \sqrt{\frac{365}{T}\sum_{t=0}^{T}(r_i - \bar{r})}, \text{ where } r_t = \frac{P_t}{P_{t-1}} - 1$$

- maximum drawdown:

$$MDD = \max_{\tau \in [0,T]}\left(\max_{t \in [0,\tau]} P_t - P_\tau\right)$$

- maximum loss duration:

$$MLD = \max_{i \in L} \frac{i}{365}, \begin{cases} L = m_j - m_k \\ V_{m_j} > V_{m_i} \\ j > k \end{cases}$$

where $m_j, m_i$ stand for numbers of days indicating consecutive local maxima of the portfolio value and $V_{m_j}, V_{m_i}$ are the corresponding values the local maxima.

- information ratios:

$$IR = \frac{aRC}{aSD}$$

$$IR2 = \frac{IR * sign(aRC) * aRC}{MD}$$

$$IR3 = \frac{aRC^3}{aSD * MDD * MLD}$$

In all performance measures' formulas $P_t$ stands for portfolio value at time $t$. To reflect the downside risk of the strategies we incorporated risk-weighted measures, since higher risk may not be desirable by the investor.

## 4 Data and Parameters Setup

### 4.1 Assets

The dataset used in this study consists of daily close prices of cryptocurrencies and US stocks downloaded respectively from Coinpaprika[1] and Yahoo Finance services. Cryptocurrency markets operate throughout the whole week, while the stock markets are closed on weekends and holidays, so the missing stock quotes for these days were filled with the last available price[2]. The initial dataset consisted of around 1000 stocks listed on the New York Stock Exchange (NYSE) and around 500 cryptocurrencies with the highest market capitalization[3]. Daily prices were collected from 2018-01-01 to 2021-10-31. We chose the stocks from NYSE, which were characterized by the overall highest market capitalization, as it allowed to ensure a variety of companies with high market shares within sectors. All stablecoins were removed from the study primarily to not cause any numerical problems during the calculations.

The initial large dataset was an input to the stage of filtering the assets based on their market capitalization. Stocks and cryptocurrencies with higher market capitalizations tend to have higher liquidity, meaning there is a larger pool of active investors and trading volume. This liquidity provides stability and reduces the risk of sudden price fluctuations, therefore we decided to use that indicator as our filtering method. Data for stocks and cryptocurrencies for each day was gathered, and then based on these values, 10 stocks and 10 cryptocurrencies with the highest market capitalization at a given date were selected and taken into the portfolio construction process. This pre-selection was done to speed up the portfolio optimization as well as select the most popular assets at a given time. To sum up, at every date of a new portfolio optimization, the assets were pre-selected based on their market capitalization and a dataset of 20 assets (10 stocks and 10 cryptocurrencies) was then passed further to the portfolio optimization step. Over the whole study horizon, this method selected 82 unique assets, including 64 cryptocurrencies and 18 stocks. The summary statistics of daily returns of both regular and crypto assets are in the Appendix to this paper incorporating the list of every ticker used in this study.

---

[1] coinpaprika.com
[2] We are aware that this influences variance and correlation, this is an obvious limitation of the study. However, removing weekend quotes of cryptocurrencies would also influence variance and correlation.
[3] Data sourced from: coinmarketcap.com

**4.2 Models' Parameters**

All strategies built in this study relied on various parameters and as many of them as possible are shared between the strategies to enable an unbiased comparison of the results. For every calculation we assumed 365 trading days per year. A window parameter corresponds to the number of past observations taken into calculation of the current variance-covariance estimator, that is how many last prices are used to calculate the matrix estimator. A rebalancing period parameter indicated the number of days passed between the two consecutive portfolios optimizations. Moreover, as described above, we assumed a joint optimization criterion for all strategies, which was chosen to be minimal variance. Table 1 summarizes the parameters common to all strategies and their possible values. Strategies based on deep learning models also share an additional parameter regarding the validation dataset – all models were trained using a validation set consisting of values from the period calculated as $window * 2$. The exact dates of training and validation periods are therefore dependent on parameters.

Table 1. Parameters and corresponding values shared by all strategies

| Parameter | Possible values |
| --- | --- |
| window | 30, 60, 90, 120 |
| rebalancing period | 30, 60, 90, 120 |
| optimization criterion | minimal variance |

*Note:* Parameter window corresponds to the number of days considered for calculation of the variance-covariance matrix estimator. Parameter rebalancing period indicates how many days passed between portfolios re-optimization. Parameters values were chosen arbitrarily.

Strategies based on the variance-covariance matrix forecast from the LSTM models were tested with numerous possible hyperparameters. The first important issue to consider was the architecture of the RNN, that is the number of hidden layers and number of units in each layer. To provide a comprehensive view on the performance of simple RNNs we used networks with a single hidden layer and two hidden layers. The possible number of units in each layer was chosen to be in range from 5 to 20. Table 2 summarizes the parameters of the investment strategies based on LSTM RNNs. Each RNN was trained with 150 epochs using ADAM optimizer with the learning rate parameter set to 0.01.

Table 2. Parameters and corresponding values used for strategies based on LSTM variance-covariance matrix forecasts

| Parameter | Possible values |
|---|---|
| Units | 5, 10, 15, 20, [5, 5], [5, 10], [10, 5], [10, 10], [15, 15], [20, 20] |
| Batch size | 8, 16 |
| Sequences length | 15, 20 |

*Note:* Numbers of units described as a single number indicate a single hidden layer, while values described with two numbers in square brackets indicate two hidden layers with corresponding number of nodes in each layer. Parameters values were chosen arbitrarily.

Probabilistic deep learning models were also tested with various sets of parameters, most of which concerned the architecture of the RNNs or the distribution used in the model. The parameters defined in Table 3 concern both DeepVAR and GPVAR architectures. All networks were trained with two hidden layers and the number of units in each layer was chosen from range between 5 and 20 by 5. Similarly as in the case of the simple LSTM models, 150 epochs were used to train the model with learning rate equal to 0.01. The scaling parameter indicated whether mean scaling was applied to inputs as in Salinas et al. (2020). Copula parameter indicated whether the model used the copula function. The low-rank flag specified whether the low-rank parametrization of the distribution variance-covariance matrix was used – otherwise a full-rank parametrization was employed.

Table 3. Parameters and corresponding values used for strategies based on probabilistic deep learning-based variance-covariance matrix forecasts

| Parameter | Possible values |
|---|---|
| Units | 5, 10, 15, 20 |
| Scaling | True, False |
| Low-Rank | True, False |
| Copula | True, False |

*Note:* Numbers of units was kept the same in both hidden layers. All parameters concern both DeepVAR and GPVAR architecture. Parameters values were chosen arbitrarily.

Throughout a single investment strategy backtest the parameters were held constant. However, on each rebalancing date a new model was created and fit to the data, since different assets could be pre-selected for that date. Therefore, duration of a backtest varied depending on the parameters – the more frequent rebalancing, the longer time required to test the strategy.

## 5 Results

To evaluate the performance of strategies based on different variance-covariance estimation methods we backtested strategies with all combinations of parameters introduced in the previous chapter. We aggregated the results across windows, rebalancing periods, and models to present them as box plots of two selected metric: information ratio and annualized returns. In such a way we were able to compare different variance-covariance estimation methods as well as check the robustness of the strategies to changes in their parameters. The numerical representation of these results was added as an appendix to this paper. Reference results based on a simple buy and hold strategy for Bitcoin and the SP500 index are added in the note to each figure.

Figure 2 presents the distributions of information ratios and annualized returns for all strategies based on a 30-day window, which means that the only last 30 observations were the input to calculation of the variance-covariance matrix estimator. Clearly all strategies were highly sensitive to the rebalancing period and less frequent portfolio re-optimizations provided better results in nearly every case. The classical variance-covariance matrix estimation methods provided the lowest information ratios across all rebalancing periods and the lowest returns in three out of four rebalancing periods. Only in the case of the shortest rebalancing window classical variance-covariance estimation methods were able to provide higher annualized returns than the probabilistic deep learning models, yet the simple LSTM models still were able to provide higher returns in most cases. Strategies based on the LSTM models outperformed all other strategies both in terms of information ratios and annualized returns. Nevertheless, for rebalancing periods above 30 days, all strategies based on probabilistic deep learning models were very stable and provided similar results, indicating that these models were quite robust to changes in their parameters and could provide good results out of the box. Nearly all strategies performed unsatisfactory providing negative or very low returns, which suggests that the input of only 30 most recent observations was not enough to properly capture the variance-covariance matrix characteristics.

Fig 2. Distributions of information ratio and annualized returns of the investment strategies based on a 30-days window.

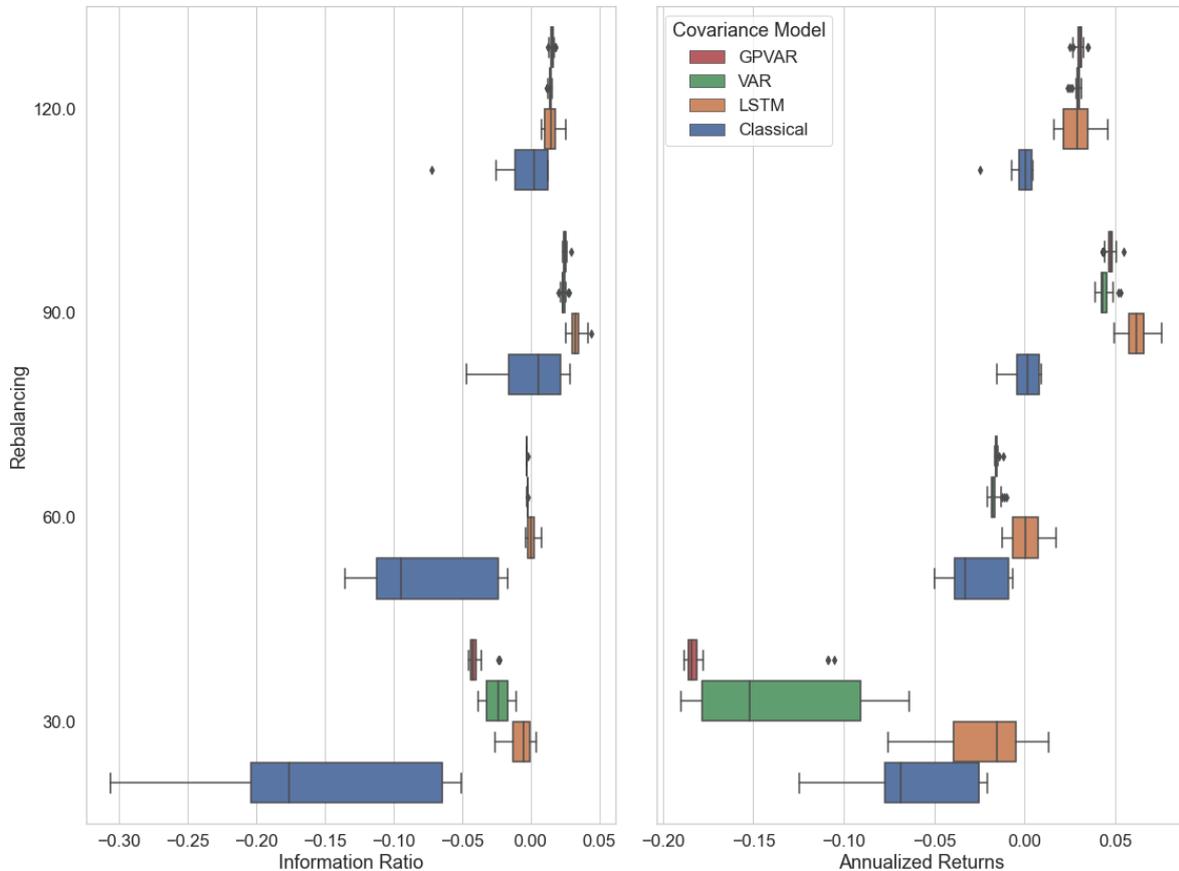

*Note:* Performance statistics were aggregated across all parameters for each variance-covariance estimation method and rebalancing period. Number of strategies in each rebalancing periods respectively was: GPVAR - 23, 32, 32, 31; VAR - 25, 32, 32, 30; LSTM - 25, 40, 39, 40; classical - 8, 8, 8, 8. The reference results are as follows B&H SP500: IR 0.699, aRC 0.15; B&H BTC: IR 0.634, aRC 0.483.

Figure 3 shows the performance metrics for all strategies based on the inputs consisting of the last 60 observations. Once again, in nearly every case, the classical variance-covariance estimation methods provided the worst results. The only exception was the 90 days rebalancing window for which all other strategies performed much worse in terms of annualized returns and information ratios. Similarly as in the case of the 30-days window, strategies with less frequent rebalancing were generally more profitable than the others. All deep learning-based methods were superior for rebalancing windows other than 90 days. Moreover, the strategies based on the LSTM models were again better than these based on the probabilistic deep learning models. The highest returns and information ratios were obtained for the re-optimizations done every 120 days. Both DeepVAR and GPVAR delivered very stable results, especially for portfolios that were rebalanced at least every 60 days.

Fig 3. Distributions of information ratio and annualized returns of the investment strategies based on a 60-days window.

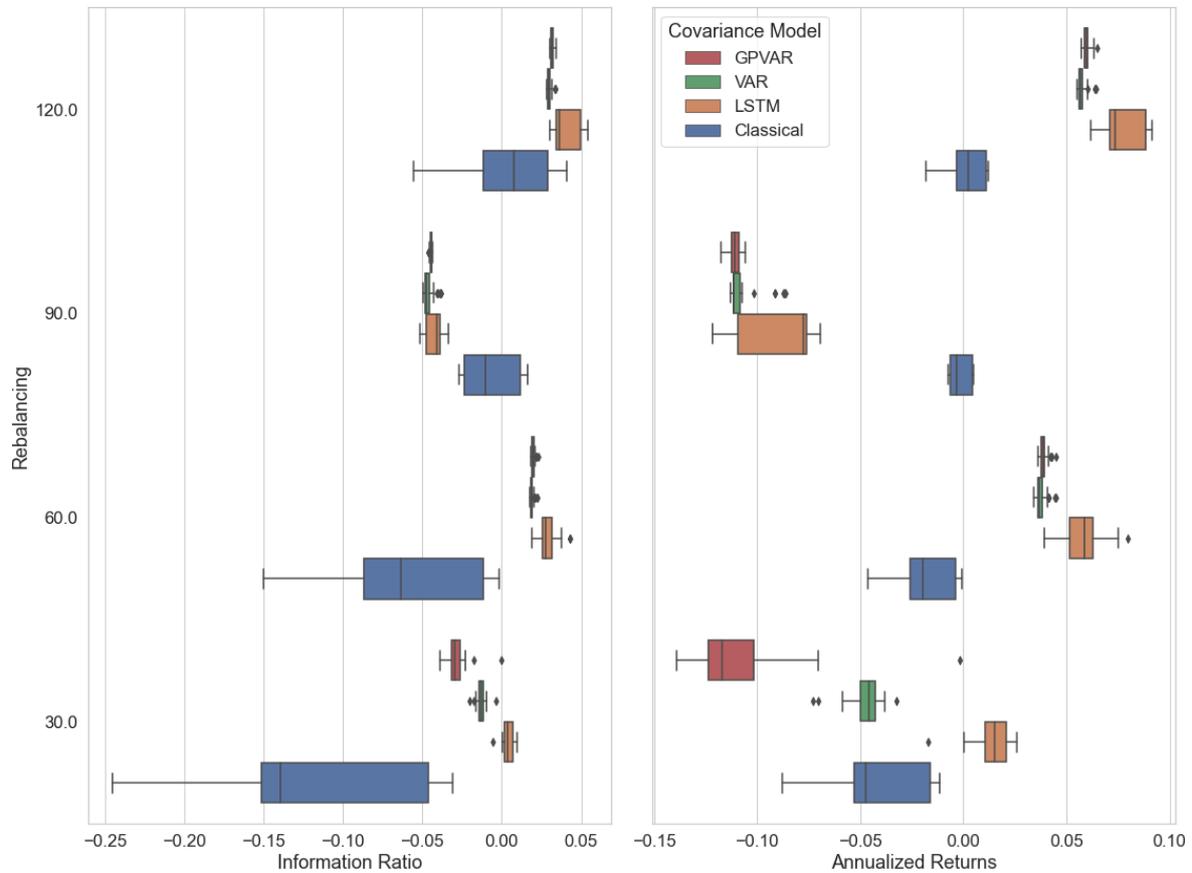

*Note:* Performance statistics were aggregated across all parameters for each variance-covariance estimation method and rebalancing period. Number of strategies in each rebalancing periods respectively was: GPVAR - 32, 32, 32, 32; VAR - 32, 32, 32, 32; LSTM - 38, 40, 40, 40; classical - 8, 8, 8, 8. The reference results are as follows B&H SP500: IR 0.699, aRC 0.15; B&H BTC: IR 0.634, aRC 0.483.

Figure 4 represents the performances of the strategies based on a 90-days observation window. In general, the strategies with portfolio rebalancing every 30, 60 or 90 days provided rather similar returns and information ratios for a given strategy. The LSTM models once again were able to produce the most profitable strategies per rebalancing frequency. The highest returns of around 10% on annual terms were reported for LSTM-based strategies with portfolios rebalanced every 120 days. Slightly worse results were obtained with DeepVAR and GPVAR models. The classical variance-covariance estimation methods provided the lowest information ratios and returns in all setups. DeepVAR and GPVAR delivered very stable results for less frequent portfolio re-optimizations. The 90-days window was the first setup when so many strategies provided positive returns and information rations. This indicates that a longer window was superior and the deep learning models need more observations to capture both short- and long-term dependencies in the variance-covariance matrices.

Fig 4. Distributions of information ratio and annualized returns of the investment strategies based on a 90-days window.

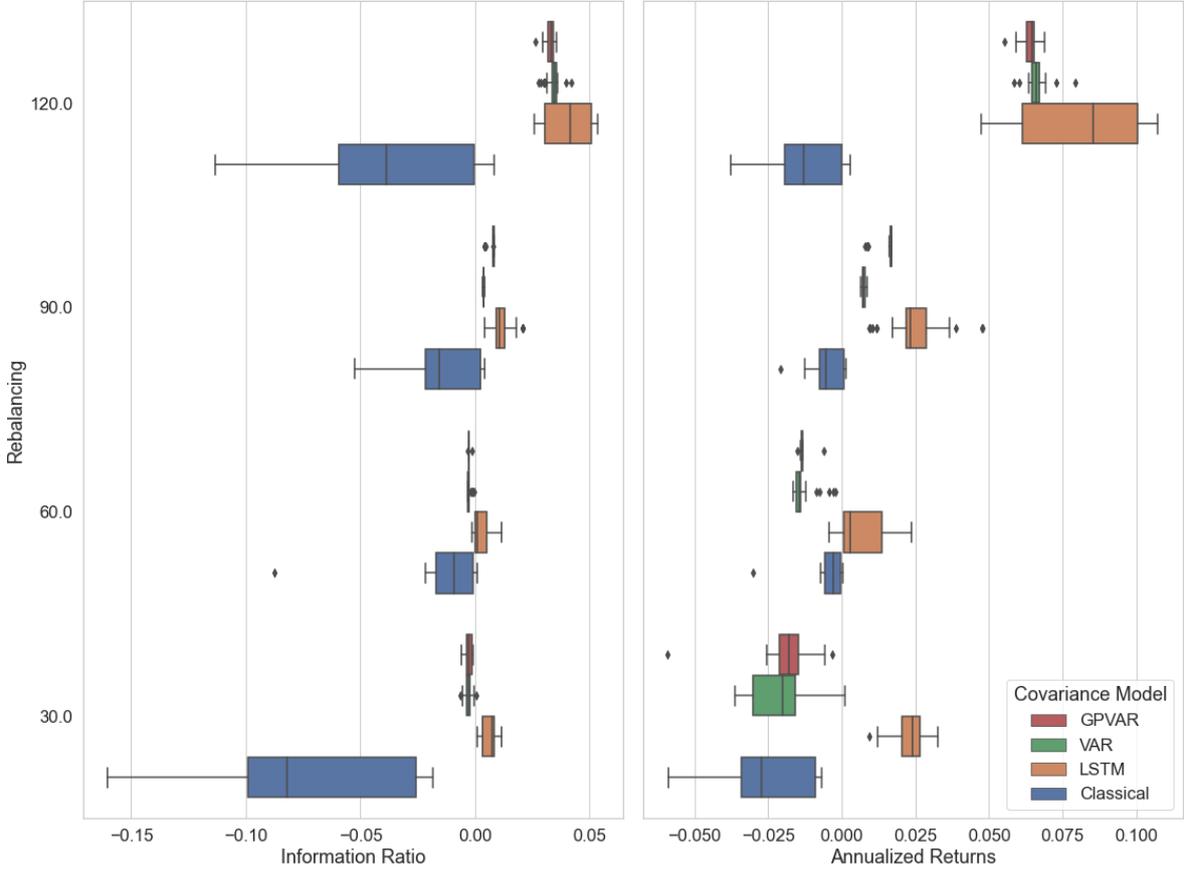

*Note:* Performance statistics were aggregated across all parameters for each variance-covariance estimation method and rebalancing period. Number of strategies in each rebalancing periods respectively was: GPVAR - 32, 32, 32, 32; VAR - 32, 32, 32, 32; LSTM - 30, 40, 40, 24; classical - 8, 8, 8, 8. The reference results are as follows B&H SP500: IR 0.699, aRC 0.15; B&H BTC: IR 0.634, aRC 0.483.

Figure 5 shows the last group of strategies, which were based on a 120-days window. All strategies based on the classical variance-covariance matrix estimation methods provided rather similar results in terms of annualized returns and information ratios, which were not satisfactory, as most of these strategies were highly unprofitable. Interestingly, the longest rebalancing period was the least profitable for the deep learning-based strategies since all of these were outperformed by the classical methods and ended up with negative returns and information ratios. For strategies with portfolios re-optimized every 30, 60 and 90 days the LSTM models outperformed all other variance-covariance estimation methods providing the highest returns and information ratios. The consistency of DeepVAR and GPVAR is observed for 60- and 90-days rebalancing periods.

Fig 5. Distributions of information ratio and annualized returns of the investment strategies based on a 120-days window.

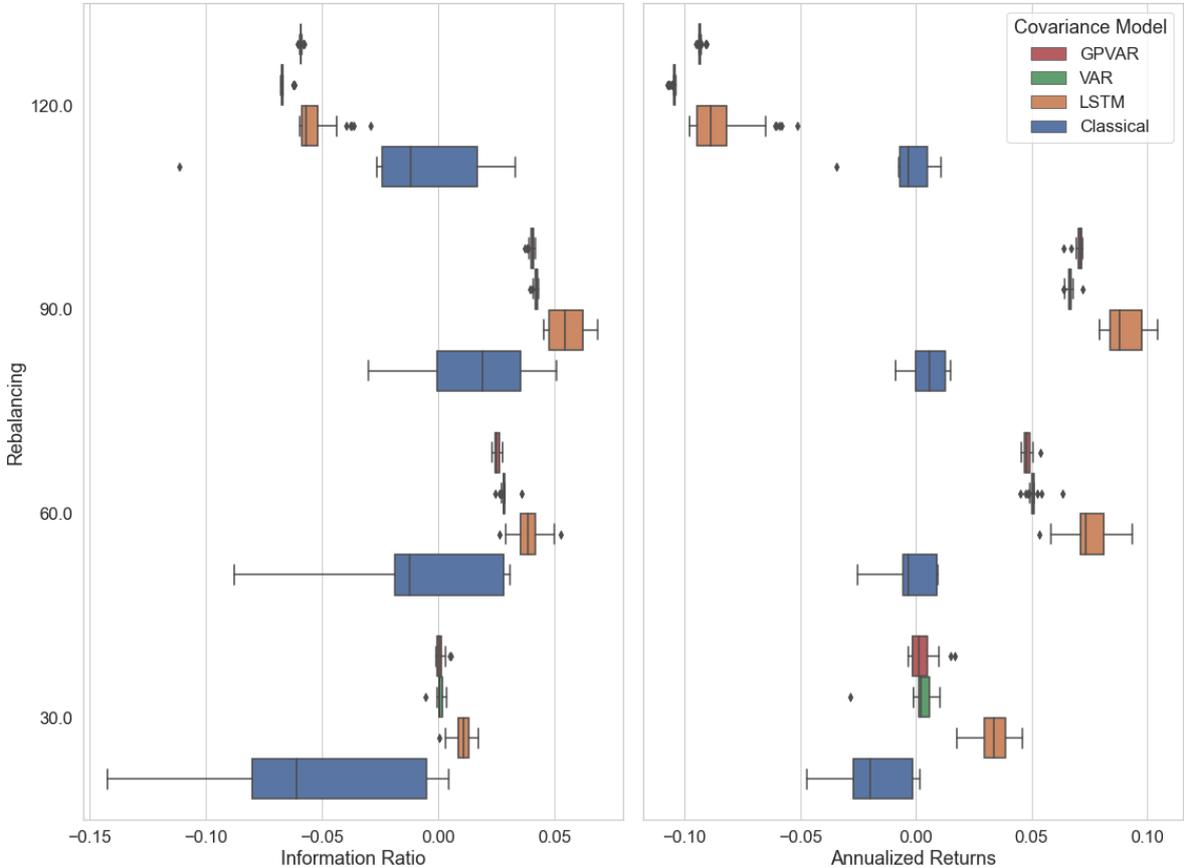

*Note:* Performance statistics were aggregated across all parameters for each variance-covariance estimation method and rebalancing period. Number of strategies in each rebalancing periods respectively was: GPVAR - 32, 32, 32, 32; VAR - 29, 32, 32, 32; LSTM - 40, 40, 40, 34; classical - 8, 8, 8, 8. The reference results are as follows B&H SP500: IR 0.699, aRC 0.15; B&H BTC: IR 0.634, aRC 0.483.

To sum up, it is clear that the strategies performance depends strongly on the assumed portfolio rebalancing and window parameter. In general, strategies with more observations considered in the variance-covariance matrix estimation calculation were better than these based on shorter windows. Moreover, a less frequent rebalancing was in multiple cases a better choice than a more frequent portfolio re-optimization. In most of the considered cases, strategies based on matrices estimated using the deep learning models performed better than these which used the classical variance-covariance estimation methods. In particular, the investment strategies that utilized forecasts from the LSTM models were generally superior over the strategies that used DeepVAR and GPVAR models. Nevertheless, the probabilistic deep learning models produced strategies which provided very similar results across all parameters indicating that these models could work very well out of the box, and it would not be necessary to spend a lot of time tuning the models' parameters. The LSTM-RNN with a proper length of historical window was able to capture the variance-covariance matrix characteristics and provide forecasts that when served as and input to the Markowitz framework provided profitable investment decisions.

The overall unsatisfactory results of the probabilistic models might be explained with a few arguments. Variances and covariances of returns used in this study constructed high-dimensional multivariate time series. Handling high-dimensional data may led to an increased model complexity, what could make it challenging for GPVAR and DeepVAR to effectively capture dependencies and estimate uncertainty in such settings. Moreover, financial returns exhibit non-stationarity and volatility clustering, meaning that the statistical properties of returns change over time and exhibit periods of high or low volatility. These characteristics made the forecasting task challenging for those models, as well as made the underlying choice of probabilistic distribution prone to errors.

**Conclusions**

The goals of this study were to present a novel variance-covariance estimation framework combining the Modern Portfolio Theory with the deep learning models and to provide an extensive comparison of various investment strategies based on portfolio optimization utilizing different variance-covariance matrix estimation methods. We introduced a detailed methodology of the variance-covariance matrix estimation with the deep learning models along with a specific description of the portfolio optimization process. Within our framework we utilized the Cholesky decomposition to model the variance-covariance matrix entrances in the multivariate deep learning models. The investment portfolios were built and backtested on almost 4 years of data, which included stocks and cryptocurrencies with the highest stock market capitalization. We used multiple classical methods which included the frequentist estimator, semi-covariance, exponentially weighted moving average, and shrinkage estimators as well as forecasting methods based on the deep learning models such as LSTM-RNN and probabilistic deep learning models: DeepVAR and GPVAR. Various combinations of parameters were used and compared to provide an exhaustive analysis of the strategies performance. The optimization objective in the Markowitz framework was minimal variance. The contributions of this paper are twofold. Firstly, the framework that we introduced may serve for further research and be used extended with different models or portfolio selection methods. Secondly, the investment strategies that we presented may constitute the ideas for market practitioners. To provide a possibility of further experimentations, the codes used to arrive at the presented results are available on our GitHub[4] repository.

Based on our framework we were able to produce strategies that provided positive returns and were profitable over the course of backtests. Nevertheless, performance of these investment schemes strongly depended on two main parameters – the length of observation window and the frequency of rebalancing. In general, a higher number of observations considered in the variance-covariance matrix estimation translated into better results, especially in the case of deep learning-based strategies. These results indicated that the deep learning models required longer historical windows to properly capture the interdependencies of the variance-covariance matrix structure. Similarly, a less frequent portfolio re-optimizations generally performed better indicating that this framework could be utilized for a long-term portfolio management. In most of the considered combinations of observation window length and rebalancing frequency, strategies based on variance-covariance matrices forecasted with the LSTM-RNN outperformed other strategies in terms of the examined performance metrics. Although, both DeepVAR and GPVAR typically achieved slightly worse results, both models were very stable across their hyperparameters, especially for longer observation windows. These results suggested that the probabilistic models are more robust to hyperparameters changes, and they could provide good results right away, without a lengthy optimization process.

The first research hypothesis concerning the performance of the deep learning-based strategies was partially rejected, as the performance of such strategies strongly depended on window and rebalancing parameters. Nevertheless, in most cases the strategies utilizing the variance-covariance matrices from the deep learning models were significantly better than the strategies exploiting the classical variance-covariance estimation methods. The second research hypothesis of this study was rejected, as the strategies employing the probabilistic deep learning models did not perform any better than the strategies with variance-covariance matrix estimation from the LSTM-RNN models.

This study presented an extensive analysis of investment strategies based on portfolio optimization utilizing both classical and deep learning models to estimate the variance-covariance matrix. We presented results combining different asset classes – stocks and cryptocurrencies – and numerous possible parameters setup. The further research extensions could incorporate larger portfolios and the dynamic financial econometrics models such as the multivariate GARCH models. Moreover, other optimization criterions for the Markowitz framework could be considered such as the Sharpe ratio optimization or maximization of returns for a given risk threshold.

---

[4] Guidance on data and codes available upon email request.

# Appendix

Table A.1. Summary statistics of daily returns of cryptocurrencies used in the study.

| Asset | Mean | Std. Dev | Min | 25% | Median | 75% | Max | Skewness | Kurtosis |
|---|---|---|---|---|---|---|---|---|---|
| NVC | 0,0114 | 0,1877 | -0,7543 | -0,0432 | -0,0006 | 0,0386 | 2,5474 | 5,1731 | 57,1950 |
| NMC | 0,0148 | 0,2937 | -0,9303 | -0,0300 | 0,0004 | 0,0331 | 7,5501 | 19,7476 | 470,4886 |
| BNB | 0,0048 | 0,0637 | -0,4380 | -0,0206 | 0,0009 | 0,0294 | 0,7010 | 1,9508 | 23,2032 |
| STEEM | 0,0008 | 0,0708 | -0,4797 | -0,0312 | -0,0011 | 0,0308 | 0,6785 | 0,8328 | 10,9348 |
| BCH | 0,0011 | 0,0658 | -0,4497 | -0,0282 | -0,0003 | 0,0266 | 0,5128 | 0,8618 | 11,1485 |
| BLK | 0,0006 | 0,0166 | -0,1365 | -0,0029 | 0,0000 | 0,0060 | 0,1352 | -0,0906 | 14,6244 |
| BTC | 0,0018 | 0,0402 | -0,3958 | -0,0152 | 0,0015 | 0,0193 | 0,1867 | -0,5149 | 8,8944 |
| APH | 0,0005 | 0,0145 | -0,1415 | -0,0029 | 0,0000 | 0,0055 | 0,1011 | -0,8309 | 14,1687 |
| XLM | 0,0017 | 0,0648 | -0,3617 | -0,0289 | -0,0009 | 0,0266 | 0,7412 | 2,0832 | 21,8284 |
| ETH | 0,0025 | 0,0516 | -0,4439 | -0,0222 | 0,0009 | 0,0289 | 0,2741 | -0,3779 | 6,3116 |
| LSK | 0,0007 | 0,0633 | -0,4422 | -0,0305 | -0,0007 | 0,0295 | 0,3444 | 0,1389 | 5,2146 |
| EOS | 0,0017 | 0,0670 | -0,4005 | -0,0248 | 0,0005 | 0,0250 | 0,5179 | 0,7947 | 8,5831 |
| EMC | 0,0030 | 0,1196 | -0,5798 | -0,0396 | -0,0047 | 0,0297 | 1,7208 | 4,7019 | 50,6566 |
| TRX | 0,0025 | 0,0716 | -0,4264 | -0,0276 | 0,0008 | 0,0289 | 1,1961 | 3,5161 | 58,1322 |
| AMP | 0,0007 | 0,0222 | -0,2382 | -0,0039 | 0,0000 | 0,0067 | 0,2264 | 0,2993 | 29,5282 |
| DOGE | 0,0065 | 0,1193 | -0,3855 | -0,0240 | -0,0011 | 0,0188 | 3,3365 | 16,7405 | 442,7789 |
| MEC | 0,0168 | 0,2302 | -0,6580 | -0,0483 | 0,0000 | 0,0467 | 3,3562 | 5,8542 | 67,0336 |
| CNC | 0,0004 | 0,0201 | -0,1501 | -0,0054 | 0,0000 | 0,0062 | 0,1560 | 0,4439 | 10,2164 |
| CLOAK | 0,0874 | 0,6348 | -0,8982 | -0,0651 | -0,0065 | 0,0626 | 8,8184 | 6,0266 | 52,9086 |
| ZET | 0,4740 | 9,3235 | -0,9961 | -0,0741 | 0,0000 | 0,0753 | 293,7987 | 27,6357 | 804,0266 |
| OMNI | 0,0055 | 0,1308 | -0,5376 | -0,0477 | -0,0016 | 0,0400 | 1,0717 | 1,9646 | 12,6693 |
| XMR | 0,0012 | 0,0535 | -0,4052 | -0,0236 | 0,0018 | 0,0273 | 0,3839 | -0,3399 | 7,8034 |
| MONA | 0,0010 | 0,0738 | -0,3292 | -0,0234 | -0,0023 | 0,0177 | 0,9541 | 4,6059 | 51,6832 |
| FCT | -0,0002 | 0,0705 | -0,3955 | -0,0351 | -0,0026 | 0,0320 | 0,4029 | 0,5002 | 4,1543 |
| NEO | 0,0014 | 0,0620 | -0,4033 | -0,0294 | 0,0002 | 0,0303 | 0,2863 | 0,0041 | 4,1679 |
| XPM | 0,0115 | 0,2006 | -0,8013 | -0,0427 | -0,0021 | 0,0360 | 3,7938 | 8,7464 | 129,5429 |
| XEM | 0,0008 | 0,0648 | -0,3217 | -0,0295 | 0,0000 | 0,0262 | 0,5431 | 0,6247 | 7,1725 |
| START | 0,0505 | 0,4686 | -0,9113 | -0,0771 | 0,0000 | 0,0717 | 10,5084 | 10,1774 | 190,4061 |
| VET | 0,0033 | 0,0783 | -0,9922 | -0,0337 | 0,0012 | 0,0359 | 0,4227 | -0,9961 | 21,9734 |
| DASH | 0,0005 | 0,0611 | -0,3981 | -0,0268 | -0,0010 | 0,0263 | 0,5668 | 1,1125 | 13,3797 |
| FTC | 0,0004 | 0,0729 | -0,3772 | -0,0341 | -0,0013 | 0,0314 | 0,3961 | 0,4482 | 4,8311 |
| DCR | 0,0017 | 0,0570 | -0,4503 | -0,0257 | 0,0007 | 0,0292 | 0,3320 | -0,0557 | 5,6111 |
| PIVX | 0,0006 | 0,0700 | -0,5112 | -0,0312 | -0,0006 | 0,0287 | 0,5409 | 0,3989 | 9,3986 |
| XRP | 0,0014 | 0,0646 | -0,4173 | -0,0243 | -0,0011 | 0,0219 | 0,5210 | 1,2067 | 12,6496 |
| DGB | 0,0025 | 0,0725 | -0,4362 | -0,0358 | -0,0027 | 0,0325 | 0,4889 | 0,6136 | 4,7882 |
| XCP | 0,0103 | 0,1863 | -0,7640 | -0,0391 | -0,0019 | 0,0349 | 3,2153 | 9,3407 | 137,7058 |
| DGC | 0,0951 | 0,8390 | -0,9550 | -0,1038 | -0,0046 | 0,1003 | 23,1309 | 16,6585 | 417,5751 |
| WAVES | 0,0026 | 0,0663 | -0,4010 | -0,0280 | 0,0011 | 0,0306 | 0,5454 | 1,0170 | 9,6781 |
| WDC | 0,0001 | 0,0270 | -0,2044 | -0,0083 | 0,0000 | 0,0094 | 0,1775 | -0,7385 | 9,8349 |
| LTC | 0,0013 | 0,0548 | -0,3808 | -0,0269 | -0,0001 | 0,0277 | 0,3372 | 0,1225 | 6,0790 |
| ADA | 0,0025 | 0,0619 | -0,4113 | -0,0289 | 0,0003 | 0,0297 | 0,3796 | 0,5279 | 4,6245 |
| PPC | 0,0010 | 0,0706 | -0,3744 | -0,0278 | -0,0016 | 0,0271 | 0,8018 | 1,9554 | 21,8106 |
| OMG | 0,0022 | 0,0717 | -0,4474 | -0,0343 | -0,0006 | 0,0330 | 0,6887 | 1,0410 | 10,5959 |
| BTG | 0,0005 | 0,0255 | -0,1879 | -0,0085 | 0,0000 | 0,0089 | 0,1611 | 0,0109 | 6,8304 |
| TRC | 0,0068 | 0,1332 | -0,7046 | -0,0487 | 0,0026 | 0,0515 | 1,5445 | 2,1203 | 21,9340 |

| Asset | Mean | Std. | Min | 25% | Median | 75% | Max | Skewness | Kurtosis |
|---|---|---|---|---|---|---|---|---|---|
| VTC | 0,0035 | 0,1078 | -0,5354 | -0,0409 | -0,0033 | 0,0338 | 1,3940 | 3,1710 | 30,3316 |
| FRC | 0,0008 | 0,0168 | -0,1435 | -0,0030 | 0,0000 | 0,0055 | 0,1288 | 0,2592 | 14,2095 |
| GLC | 0,0364 | 0,6089 | -0,9296 | -0,0329 | 0,0000 | 0,0253 | 15,5224 | 18,4268 | 393,8562 |
| IXC | 0,1834 | 2,8447 | -0,9924 | -0,0625 | 0,0000 | 0,0576 | 80,8915 | 23,7855 | 609,3447 |
| QRK | 0,0298 | 0,3714 | -0,9146 | -0,0350 | -0,0004 | 0,0301 | 9,9635 | 15,3267 | 374,9162 |
| TRMB | 0,0007 | 0,0195 | -0,1860 | -0,0035 | 0,0000 | 0,0070 | 0,1881 | 0,0640 | 21,3127 |
| LUNA | 0,1315 | 1,8073 | -0,9827 | -0,0631 | -0,0037 | 0,0568 | 60,4776 | 27,4960 | 894,5993 |
| XTZ | 0,0025 | 0,0685 | -0,4640 | -0,0323 | 0,0003 | 0,0341 | 0,2944 | -0,0115 | 4,1450 |
| PLU | 0,0060 | 0,1273 | -0,5389 | -0,0599 | -0,0043 | 0,0552 | 1,1802 | 2,1756 | 14,7743 |
| LINK | 0,0056 | 0,0767 | -0,4927 | -0,0354 | 0,0014 | 0,0398 | 0,6099 | 0,8994 | 8,2981 |
| ZEC | 0,0009 | 0,0593 | -0,4098 | -0,0284 | -0,0006 | 0,0314 | 0,2979 | -0,0716 | 4,6661 |
| REP | 0,0013 | 0,0668 | -0,4199 | -0,0302 | 0,0006 | 0,0286 | 0,6391 | 1,3054 | 13,5917 |
| USNBT | 0,0223 | 0,2891 | -0,8794 | -0,0503 | -0,0005 | 0,0478 | 5,3951 | 9,9010 | 157,4541 |
| VRC | 0,0102 | 0,1938 | -0,5908 | -0,0551 | -0,0048 | 0,0517 | 4,0747 | 9,1211 | 165,0179 |
| MIOTA | 0,0012 | 0,0623 | -0,4258 | -0,0301 | -0,0005 | 0,0320 | 0,3595 | 0,0836 | 5,2183 |
| VERI | 0,0130 | 0,2007 | -0,7488 | -0,0601 | -0,0046 | 0,0582 | 2,4168 | 5,5865 | 57,3260 |
| CLR | 0,0007 | 0,0383 | -0,5253 | -0,0096 | 0,0000 | 0,0098 | 0,3319 | -0,6294 | 33,7649 |
| ETC | 0,0023 | 0,0639 | -0,4208 | -0,0233 | 0,0005 | 0,0251 | 0,4442 | 0,8263 | 9,3156 |
| BTS | 0,0006 | 0,0752 | -0,3909 | -0,0277 | -0,0008 | 0,0291 | 1,1661 | 3,4990 | 51,7868 |

*Note:* Asset names correspond to tickers from cryptocurrency exchanges.

Table A.2. Summary statistics of daily returns of stocks used in the study.

| Asset | Mean | Std. | Min | 25% | Media | 75% | Max | Skewnes | Kurtosi |
|---|---|---|---|---|---|---|---|---|---|
| JNJ | 0,000 | 0,0115 | - | - | 0,00 | 0,003 | 0,08 | -0,482 | 15,098 |
| UNH | 0,000 | 0,0164 | - | - | 0,00 | 0,005 | 0,12 | -0,086 | 19,784 |
| AAPL | 0,001 | 0,0174 | - | - | 0,00 | 0,006 | 0,12 | -0,048 | 9,639 |
| V | 0,000 | 0,0152 | - | - | 0,00 | 0,005 | 0,13 | 0,149 | 16,369 |
| CVX | 0,000 | 0,0193 | - | - | 0,00 | 0,004 | 0,22 | -0,305 | 36,395 |
| PG | 0,000 | 0,0116 | - | - | 0,00 | 0,003 | 0,12 | 0,666 | 20,924 |
| MSFT | 0,001 | 0,0158 | - | - | 0,00 | 0,006 | 0,14 | 0,032 | 14,866 |
| BAC | 0,000 | 0,0192 | - | - | 0,00 | 0,006 | 0,17 | 0,500 | 18,827 |
| GOOG | 0,000 | 0,0154 | - | - | 0,00 | 0,005 | 0,10 | -0,018 | 8,854 |
| WMT | 0,000 | 0,0120 | - | - | 0,00 | 0,003 | 0,11 | 0,869 | 22,743 |
| WFC | 0,000 | 0,0201 | - | - | 0,00 | 0,005 | 0,14 | -0,017 | 13,150 |
| TSLA | 0,002 | 0,0341 | - | - | 0,00 | 0,012 | 0,19 | 0,446 | 7,949 |
| AMZ | 0,000 | 0,0164 | - | - | 0,00 | 0,006 | 0,09 | 0,054 | 5,164 |
| NVDA | 0,001 | 0,0250 | - | - | 0,00 | 0,010 | 0,17 | -0,566 | 8,752 |
| JPM | 0,000 | 0,0173 | - | - | 0,00 | 0,004 | 0,18 | 0,525 | 22,369 |
| XOM | 0,000 | 0,0178 | - | - | 0,00 | 0,005 | 0,12 | 0,185 | 10,352 |
| FB | 0,000 | 0,0190 | - | - | 0,00 | 0,007 | 0,10 | -0,684 | 13,697 |
| BRK | 0,000 | 0,0124 | - | - | 0,00 | 0,004 | 0,11 | -0,061 | 18,183 |

*Note:* Asset names correspond to tickers from New York Stock Exchange.

Table A.3. Summary statistics of annualized returns of all investment strategies.

| Window | Rebalancing | Covariance Model | Mean | Std. Dev | Min | 25% | 50% | 75% | Max | N |
|---|---|---|---|---|---|---|---|---|---|---|
| 30 | 30 | Classical | -0,060 | 0,036 | -0,124 | -0,077 | -0,068 | -0,025 | -0,021 | 1369 |
| | | GPVAR | -0,174 | 0,026 | -0,189 | -0,186 | -0,184 | -0,181 | -0,106 | |
| | | LSTM | -0,021 | 0,022 | -0,076 | -0,039 | -0,015 | -0,005 | 0,013 | |
| | | VAR | -0,135 | 0,046 | -0,190 | -0,178 | -0,152 | -0,091 | -0,064 | |
| | 60 | Classical | -0,027 | 0,017 | -0,050 | -0,039 | -0,033 | -0,009 | -0,007 | |
| | | GPVAR | -0,016 | 0,001 | -0,017 | -0,016 | -0,016 | -0,015 | -0,012 | |
| | | lstm | 0,001 | 0,008 | -0,013 | -0,007 | 0,000 | 0,007 | 0,017 | |
| | | VAR | -0,017 | 0,003 | -0,021 | -0,019 | -0,018 | -0,016 | -0,010 | |
| | 90 | Classical | 0,000 | 0,009 | -0,015 | -0,004 | 0,002 | 0,008 | 0,009 | |
| | | GPVAR | 0,047 | 0,002 | 0,043 | 0,047 | 0,048 | 0,049 | 0,055 | |
| | | LSTM | 0,062 | 0,007 | 0,049 | 0,057 | 0,062 | 0,066 | 0,076 | |
| | | VAR | 0,044 | 0,003 | 0,039 | 0,042 | 0,043 | 0,045 | 0,053 | |
| | 120 | Classical | -0,002 | 0,010 | -0,025 | -0,003 | 0,001 | 0,004 | 0,005 | |
| | | GPVAR | 0,030 | 0,002 | 0,025 | 0,029 | 0,031 | 0,031 | 0,035 | |
| | | LSTM | 0,029 | 0,009 | 0,016 | 0,021 | 0,029 | 0,035 | 0,046 | |
| | | VAR | 0,029 | 0,002 | 0,024 | 0,029 | 0,029 | 0,030 | 0,031 | |
| 60 | 30 | Classical | -0,041 | 0,027 | -0,088 | -0,053 | -0,047 | -0,016 | -0,011 | 1339 |
| | | GPVAR | -0,110 | 0,025 | -0,139 | -0,124 | -0,117 | -0,102 | -0,002 | |
| | | LSTM | 0,015 | 0,008 | -0,017 | 0,011 | 0,015 | 0,021 | 0,026 | |
| | | VAR | -0,047 | 0,008 | -0,073 | -0,050 | -0,046 | -0,043 | -0,032 | |
| | 60 | Classical | -0,018 | 0,016 | -0,046 | -0,026 | -0,020 | -0,004 | -0,001 | |
| | | GPVAR | 0,039 | 0,002 | 0,036 | 0,038 | 0,039 | 0,039 | 0,045 | |
| | | LSTM | 0,058 | 0,009 | 0,039 | 0,052 | 0,059 | 0,063 | 0,080 | |
| | | VAR | 0,038 | 0,002 | 0,034 | 0,036 | 0,037 | 0,038 | 0,045 | |
| | 90 | Classical | -0,002 | 0,006 | -0,007 | -0,007 | -0,003 | 0,004 | 0,005 | |
| | | GPVAR | -0,111 | 0,003 | -0,118 | -0,113 | -0,111 | -0,109 | -0,106 | |
| | | LSTM | -0,090 | 0,017 | -0,122 | -0,109 | -0,078 | -0,076 | -0,069 | |
| | | VAR | -0,107 | 0,008 | -0,113 | -0,112 | -0,111 | -0,108 | -0,086 | |
| | 120 | Classical | 0,002 | 0,011 | -0,018 | -0,003 | 0,003 | 0,011 | 0,012 | |
| | | GPVAR | 0,060 | 0,002 | 0,057 | 0,059 | 0,060 | 0,060 | 0,065 | |
| | | LSTM | 0,077 | 0,010 | 0,062 | 0,071 | 0,074 | 0,088 | 0,091 | |
| | | VAR | 0,058 | 0,002 | 0,055 | 0,056 | 0,057 | 0,058 | 0,064 | |
| 90 | 30 | Classical | -0,027 | 0,019 | -0,059 | -0,034 | -0,027 | -0,009 | -0,007 | 1309 |
| | | GPVAR | -0,018 | 0,009 | -0,059 | -0,021 | -0,018 | -0,015 | -0,003 | |
| | | LSTM | 0,023 | 0,005 | 0,009 | 0,020 | 0,024 | 0,027 | 0,033 | |
| | | VAR | -0,022 | 0,009 | -0,036 | -0,030 | -0,020 | -0,016 | 0,001 | |
| | 60 | Classical | -0,006 | 0,010 | -0,030 | -0,006 | -0,003 | 0,000 | 0,000 | |
| | | GPVAR | -0,013 | 0,001 | -0,015 | -0,014 | -0,014 | -0,013 | -0,006 | |
| | | LSTM | 0,007 | 0,008 | -0,004 | 0,001 | 0,003 | 0,014 | 0,024 | |
| | | VAR | -0,013 | 0,004 | -0,016 | -0,015 | -0,015 | -0,014 | -0,002 | |
| | 90 | Classical | -0,006 | 0,008 | -0,021 | -0,008 | -0,005 | 0,001 | 0,002 | |
| | | GPVAR | 0,015 | 0,003 | 0,008 | 0,016 | 0,017 | 0,017 | 0,017 | |
| | | LSTM | 0,025 | 0,009 | 0,009 | 0,022 | 0,023 | 0,029 | 0,048 | |
| | | VAR | 0,007 | 0,001 | 0,006 | 0,007 | 0,007 | 0,008 | 0,009 | |
| | 120 | Classical | -0,012 | 0,014 | -0,037 | -0,020 | -0,013 | 0,000 | 0,003 | |
| | | GPVAR | 0,064 | 0,003 | 0,055 | 0,063 | 0,064 | 0,065 | 0,069 | |
| | | LSTM | 0,080 | 0,020 | 0,047 | 0,061 | 0,085 | 0,100 | 0,107 | |
| | | VAR | 0,066 | 0,003 | 0,058 | 0,065 | 0,066 | 0,067 | 0,079 | |
| 120 | 30 | Classical | -0,019 | 0,019 | -0,047 | -0,027 | -0,020 | -0,002 | 0,002 | 1279 |
| | | GPVAR | 0,003 | 0,005 | -0,003 | -0,001 | 0,001 | 0,005 | 0,017 | |
| | | LSTM | 0,034 | 0,007 | 0,018 | 0,030 | 0,034 | 0,039 | 0,046 | |
| | | VAR | 0,003 | 0,007 | -0,028 | 0,001 | 0,002 | 0,006 | 0,011 | |
| | 60 | Classical | -0,002 | 0,012 | -0,025 | -0,005 | -0,003 | 0,009 | 0,010 | |
| | | GPVAR | 0,048 | 0,002 | 0,046 | 0,047 | 0,048 | 0,049 | 0,054 | |
| | | LSTM | 0,076 | 0,010 | 0,054 | 0,071 | 0,073 | 0,081 | 0,094 | |
| | | VAR | 0,051 | 0,003 | 0,045 | 0,050 | 0,051 | 0,051 | 0,063 | |
| | 90 | Classical | 0,005 | 0,008 | -0,009 | 0,000 | 0,006 | 0,013 | 0,015 | |
| | | GPVAR | 0,071 | 0,002 | 0,064 | 0,070 | 0,071 | 0,071 | 0,072 | |
| | | LSTM | 0,091 | 0,008 | 0,080 | 0,084 | 0,088 | 0,098 | 0,105 | |
| | | VAR | 0,067 | 0,001 | 0,064 | 0,066 | 0,066 | 0,067 | 0,072 | |
| | 120 | Classical | -0,004 | 0,014 | -0,034 | -0,007 | -0,003 | 0,005 | 0,011 | |

| | | | | | | | | | |
|---|---|---|---|---|---|---|---|---|---|
| | | GPVAR | -0,093 | 0,001 | -0,095 | -0,094 | -0,093 | -0,093 | -0,091 |
| | | LSTM | -0,085 | 0,013 | -0,098 | -0,095 | -0,089 | -0,082 | -0,051 |
| | | VAR | -0,105 | 0,001 | -0,107 | -0,105 | -0,104 | -0,104 | -0,104 |

*Note:* Presented results were aggregated over all combinations of parameters across window, rebalancing, and variance-covariance estimation method. Parameters specifications that did not converge were removed, since such strategies could not be backtested. *N* is the number of days included in the strategy.

Table A.4. Summary statistics of information ratios of all investment strategies.

| Window | Rebalancing | Covariance Model | Mean | Std. Dev | Min | 25% | 50% | 75% | Max | N |
|---|---|---|---|---|---|---|---|---|---|---|
| 30 | 30 | Classical | -0,155 | 0,091 | -0,306 | -0,204 | -0,176 | -0,064 | -0,051 | 1369 |
| | | GPVAR | -0,039 | 0,007 | -0,045 | -0,044 | -0,042 | -0,040 | -0,023 | |
| | | LSTM | -0,007 | 0,008 | -0,026 | -0,013 | -0,005 | 0,000 | 0,004 | |
| | | VAR | -0,024 | 0,009 | -0,038 | -0,032 | -0,024 | -0,016 | -0,010 | |
| | 60 | Classical | -0,076 | 0,049 | -0,135 | -0,112 | -0,095 | -0,024 | -0,016 | |
| | | GPVAR | -0,003 | 0,000 | -0,003 | -0,003 | -0,003 | -0,003 | -0,002 | |
| | | LSTM | 0,001 | 0,003 | -0,003 | -0,002 | 0,000 | 0,003 | 0,008 | |
| | | VAR | -0,002 | 0,000 | -0,003 | -0,002 | -0,002 | -0,002 | -0,002 | |
| | 90 | Classical | 0,000 | 0,027 | -0,047 | -0,016 | 0,005 | 0,022 | 0,028 | |
| | | GPVAR | 0,025 | 0,001 | 0,023 | 0,024 | 0,025 | 0,025 | 0,030 | |
| | | LSTM | 0,033 | 0,004 | 0,026 | 0,030 | 0,033 | 0,035 | 0,044 | |
| | | VAR | 0,024 | 0,002 | 0,020 | 0,023 | 0,024 | 0,025 | 0,028 | |
| | 120 | Classical | -0,008 | 0,029 | -0,072 | -0,011 | 0,002 | 0,012 | 0,012 | |
| | | GPVAR | 0,016 | 0,001 | 0,012 | 0,015 | 0,016 | 0,016 | 0,018 | |
| | | LSTM | 0,015 | 0,005 | 0,008 | 0,010 | 0,015 | 0,018 | 0,026 | |
| | | VAR | 0,014 | 0,001 | 0,011 | 0,014 | 0,014 | 0,015 | 0,016 | |
| 60 | 30 | Classical | -0,118 | 0,075 | -0,245 | -0,151 | -0,139 | -0,046 | -0,031 | 1339 |
| | | GPVAR | -0,029 | 0,007 | -0,039 | -0,032 | -0,030 | -0,026 | 0,000 | |
| | | LSTM | 0,004 | 0,003 | -0,005 | 0,002 | 0,004 | 0,007 | 0,010 | |
| | | VAR | -0,013 | 0,003 | -0,020 | -0,014 | -0,013 | -0,012 | -0,003 | |
| | 60 | Classical | -0,059 | 0,052 | -0,150 | -0,087 | -0,064 | -0,012 | -0,002 | |
| | | GPVAR | 0,020 | 0,001 | 0,018 | 0,019 | 0,020 | 0,020 | 0,023 | |
| | | LSTM | 0,029 | 0,006 | 0,019 | 0,026 | 0,028 | 0,032 | 0,043 | |
| | | VAR | 0,019 | 0,001 | 0,018 | 0,018 | 0,019 | 0,019 | 0,023 | |
| | 90 | Classical | -0,007 | 0,018 | -0,027 | -0,024 | -0,011 | 0,012 | 0,016 | |
| | | GPVAR | -0,044 | 0,001 | -0,046 | -0,045 | -0,044 | -0,044 | -0,044 | |
| | | LSTM | -0,043 | 0,005 | -0,052 | -0,047 | -0,041 | -0,039 | -0,034 | |
| | | VAR | -0,046 | 0,003 | -0,049 | -0,048 | -0,048 | -0,045 | -0,038 | |
| | 120 | Classical | 0,004 | 0,032 | -0,055 | -0,011 | 0,008 | 0,029 | 0,041 | |
| | | GPVAR | 0,032 | 0,001 | 0,030 | 0,031 | 0,032 | 0,033 | 0,035 | |
| | | LSTM | 0,041 | 0,008 | 0,030 | 0,035 | 0,036 | 0,049 | 0,054 | |
| | | VAR | 0,030 | 0,001 | 0,029 | 0,029 | 0,030 | 0,031 | 0,033 | |
| 90 | 30 | Classical | -0,076 | 0,052 | -0,160 | -0,099 | -0,082 | -0,026 | -0,019 | 1309 |
| | | GPVAR | -0,003 | 0,001 | -0,006 | -0,004 | -0,003 | -0,001 | -0,001 | |
| | | LSTM | 0,006 | 0,003 | 0,001 | 0,003 | 0,007 | 0,008 | 0,011 | |
| | | VAR | -0,003 | 0,002 | -0,007 | -0,004 | -0,003 | -0,002 | 0,000 | |
| | 60 | Classical | -0,018 | 0,029 | -0,088 | -0,017 | -0,009 | -0,001 | 0,001 | |
| | | GPVAR | -0,003 | 0,000 | -0,003 | -0,003 | -0,003 | -0,003 | -0,001 | |
| | | LSTM | 0,003 | 0,004 | -0,001 | 0,000 | 0,001 | 0,005 | 0,011 | |
| | | VAR | -0,003 | 0,001 | -0,003 | -0,003 | -0,003 | -0,003 | -0,001 | |
| | 90 | Classical | -0,016 | 0,020 | -0,053 | -0,022 | -0,016 | 0,002 | 0,004 | |
| | | GPVAR | 0,007 | 0,001 | 0,004 | 0,008 | 0,008 | 0,008 | 0,008 | |
| | | LSTM | 0,011 | 0,004 | 0,004 | 0,009 | 0,011 | 0,013 | 0,021 | |
| | | VAR | 0,004 | 0,000 | 0,003 | 0,004 | 0,004 | 0,004 | 0,004 | |
| | 120 | Classical | -0,037 | 0,043 | -0,113 | -0,060 | -0,039 | 0,000 | 0,008 | |
| | | GPVAR | 0,033 | 0,002 | 0,026 | 0,032 | 0,033 | 0,034 | 0,036 | |
| | | LSTM | 0,040 | 0,010 | 0,026 | 0,030 | 0,042 | 0,051 | 0,054 | |
| | | VAR | 0,034 | 0,003 | 0,028 | 0,034 | 0,035 | 0,035 | 0,042 | |
| 120 | 30 | Classical | -0,056 | 0,055 | -0,142 | -0,080 | -0,061 | -0,005 | 0,005 | 1279 |
| | | GPVAR | 0,001 | 0,002 | -0,001 | 0,000 | 0,000 | 0,001 | 0,006 | |
| | | LSTM | 0,011 | 0,004 | 0,000 | 0,009 | 0,011 | 0,013 | 0,017 | |
| | | VAR | 0,001 | 0,002 | -0,005 | 0,000 | 0,001 | 0,002 | 0,003 | |
| | 60 | Classical | -0,008 | 0,039 | -0,088 | -0,019 | -0,012 | 0,028 | 0,031 | |
| | | GPVAR | 0,025 | 0,001 | 0,023 | 0,025 | 0,025 | 0,026 | 0,028 | |
| | | LSTM | 0,039 | 0,005 | 0,026 | 0,036 | 0,039 | 0,042 | 0,052 | |
| | | VAR | 0,028 | 0,002 | 0,025 | 0,028 | 0,028 | 0,029 | 0,036 | |
| | 90 | Classical | 0,016 | 0,027 | -0,030 | 0,000 | 0,019 | 0,036 | 0,051 | |
| | | GPVAR | 0,040 | 0,001 | 0,037 | 0,040 | 0,041 | 0,041 | 0,042 | |
| | | LSTM | 0,056 | 0,008 | 0,046 | 0,048 | 0,055 | 0,062 | 0,069 | |
| | | VAR | 0,042 | 0,001 | 0,039 | 0,042 | 0,042 | 0,043 | 0,043 | |
| | 120 | Classical | -0,013 | 0,046 | -0,111 | -0,024 | -0,012 | 0,017 | 0,033 | |
| | | GPVAR | -0,059 | 0,000 | -0,060 | -0,059 | -0,059 | -0,059 | -0,058 | |
| | | LSTM | -0,053 | 0,008 | -0,060 | -0,058 | -0,057 | -0,052 | -0,029 | |
| | | VAR | -0,067 | 0,002 | -0,068 | -0,067 | -0,067 | -0,067 | -0,062 | |

*Note:* Presented results were aggregated over all combinations of parameters across window, rebalancing, and variance-covariance estimation method. Parameters specifications that did not converge were removed, since such strategies could not be backtested. $N$ is the number of days included in the strategy.

Table A.5. Strategies performance metrics and parameters for 3 best and 3 worst strategies based on a 30-days window.

| 30 Rebalancing | Covariance Model | Strategy Rank | aRC | aSD | MD | MLD | IR | IR2 | IR3 | Batch Size | Length | Cells | Scaling | Copula | Low Rank |
|---|---|---|---|---|---|---|---|---|---|---|---|---|---|---|---|
| 30 | Cl as | To p | -0,021 | 0,405 | 0,436 | 3,510 | -0,051 | -0,002 | 0,000 | | | | | | |

| | | | | | | | | | | | | | | |
|---|---|---|---|---|---|---|---|---|---|---|---|---|---|---|
| | | | -0,021 | 0,405 | 0,436 | 3,510 | -0,051 | -0,002 | 0,000 | | | | | |
| | | | -0,027 | 0,389 | 0,446 | 3,510 | -0,069 | -0,004 | 0,000 | | | | | |
| | | | -0,077 | 0,364 | 0,426 | 3,510 | -0,211 | -0,038 | -0,001 | | | | | |
| | | Bottom | -0,078 | 0,388 | 0,464 | 3,510 | -0,201 | -0,034 | -0,001 | | | | | |
| | | | -0,124 | 0,406 | 0,507 | 3,268 | -0,306 | -0,075 | -0,003 | | | | | |
| | | Top | 0,013 | 3,560 | 0,898 | 2,784 | 0,004 | 0,000 | 0,000 | 16 | 20 | [15] | | |
| | | | 0,006 | 2,447 | 0,830 | 3,510 | 0,002 | 0,000 | 0,000 | 16 | 20 | [5] | | |
| | LSTM | | 0,003 | 2,799 | 0,831 | 3,510 | 0,001 | 0,000 | 0,000 | 16 | 20 | [5, 5] | | |
| | | Bottom | -0,050 | 3,266 | 0,865 | 3,014 | -0,015 | -0,001 | 0,000 | 8 | 15 | [20] | | |
| | | | -0,052 | 2,694 | 0,895 | 2,973 | -0,019 | -0,001 | 0,000 | 8 | 15 | [20, 20] | | |
| | | | -0,076 | 2,898 | 0,896 | 2,784 | -0,026 | -0,002 | 0,000 | 8 | 20 | [20, 20] | | |
| | | Top | -0,106 | 4,456 | 0,989 | 3,014 | -0,024 | -0,003 | 0,000 | 16 | | 5 | False | False | True |
| | | | -0,109 | 4,749 | 0,989 | 3,014 | -0,023 | -0,003 | 0,000 | 16 | | 20 | True | True | True |
| | GPVAR | | -0,109 | 4,755 | 0,988 | 3,014 | -0,023 | -0,003 | 0,000 | 16 | | 5 | True | False | True |
| | | Bottom | -0,187 | 4,423 | 0,988 | 3,014 | -0,042 | -0,008 | 0,000 | 16 | | 20 | True | True | False |
| | | | -0,188 | 5,203 | 0,989 | 3,014 | -0,036 | -0,007 | 0,000 | 16 | | 15 | True | False | False |
| | | | -0,189 | 4,735 | 0,989 | 3,014 | -0,040 | -0,008 | 0,000 | 16 | | 15 | False | True | True |
| | | Top | -0,064 | 6,147 | 0,990 | 3,014 | -0,010 | -0,001 | 0,000 | 16 | | 5 | False | False | False |
| | | | -0,078 | 4,433 | 0,985 | 3,014 | -0,018 | -0,001 | 0,000 | 16 | | 20 | True | False | True |
| | VAR | | -0,086 | 4,529 | 0,986 | 2,792 | -0,019 | -0,002 | 0,000 | 16 | | 10 | False | True | True |
| | | Bottom | -0,187 | 5,780 | 0,988 | 3,014 | -0,032 | -0,006 | 0,000 | 16 | | 15 | True | True | True |
| | | | -0,189 | 5,685 | 0,989 | 3,014 | -0,033 | -0,006 | 0,000 | 16 | | 15 | True | False | True |
| | | | -0,190 | 6,218 | 0,990 | 3,014 | -0,031 | -0,006 | 0,000 | 16 | | 10 | False | False | True |
| | | Top | -0,007 | 0,395 | 0,399 | 1,964 | -0,016 | 0,000 | 0,000 | | | | | |
| | | | -0,007 | 0,395 | 0,399 | 1,964 | -0,016 | 0,000 | 0,000 | | | | | |
| | Classical | | -0,010 | 0,376 | 0,403 | 1,455 | -0,026 | -0,001 | 0,000 | | | | | |
| | | Bottom | -0,038 | 0,330 | 0,350 | 1,504 | -0,114 | -0,012 | 0,000 | | | | | |
| | | | -0,041 | 0,370 | 0,421 | 2,455 | -0,112 | -0,011 | 0,000 | | | | | |
| | | | -0,050 | 0,370 | 0,411 | 1,504 | -0,135 | -0,016 | -0,001 | | | | | |
| | | Top | 0,017 | 2,115 | 0,891 | 1,989 | 0,008 | 0,000 | 0,000 | 16 | 20 | [10, 5] | | |
| | | | 0,017 | 2,128 | 0,891 | 1,989 | 0,008 | 0,000 | 0,000 | 16 | 20 | [10, 5] | | |
| | LSTM | | 0,017 | 2,134 | 0,891 | 1,989 | 0,008 | 0,000 | 0,000 | 16 | 20 | [5] | | |
| 60 | | Bottom | -0,009 | 3,548 | 0,920 | 1,992 | -0,002 | 0,000 | 0,000 | 16 | 20 | [20, 20] | | |
| | | | -0,010 | 3,756 | 0,920 | 1,989 | -0,003 | 0,000 | 0,000 | 16 | 20 | [15] | | |
| | | | -0,013 | 3,771 | 0,920 | 1,992 | -0,003 | 0,000 | 0,000 | 16 | 15 | [20] | | |
| | | Top | -0,012 | 4,782 | 0,944 | 2,690 | -0,002 | 0,000 | 0,000 | 16 | | 5 | False | False | True |
| | | | -0,014 | 5,186 | 0,945 | 2,682 | -0,003 | 0,000 | 0,000 | 16 | | 20 | True | True | True |
| | GPVAR | | -0,015 | 5,502 | 0,945 | 2,682 | -0,003 | 0,000 | 0,000 | 16 | | 15 | True | False | False |
| | | Bottom | -0,017 | 6,105 | 0,947 | 2,682 | -0,003 | 0,000 | 0,000 | 16 | | 20 | False | True | False |
| | | | -0,017 | 5,928 | 0,949 | 2,682 | -0,003 | 0,000 | 0,000 | 16 | | 5 | False | True | True |
| | | | -0,017 | 5,954 | 0,948 | 1,992 | -0,003 | 0,000 | 0,000 | 16 | | 5 | False | False | False |
| | | Top | -0,010 | 5,435 | 0,927 | 2,811 | -0,002 | 0,000 | 0,000 | 16 | | 5 | True | True | True |
| | | | -0,011 | 5,442 | 0,944 | 1,989 | -0,002 | 0,000 | 0,000 | 16 | | 5 | False | True | False |
| | VAR | | -0,013 | 5,879 | 0,946 | 1,989 | -0,002 | 0,000 | 0,000 | 16 | | 20 | False | True | False |
| | | Bottom | -0,020 | 8,998 | 0,950 | 1,992 | -0,002 | 0,000 | 0,000 | 16 | | 5 | False | True | True |
| | | | -0,020 | 9,047 | 0,955 | 1,992 | -0,002 | 0,000 | 0,000 | 16 | | 5 | True | False | True |
| | | | -0,021 | 9,139 | 0,955 | 1,992 | -0,002 | 0,000 | 0,000 | 16 | | 20 | True | True | False |
| | | Top | 0,009 | 0,328 | 0,333 | 3,510 | 0,028 | 0,001 | 0,000 | | | | | |
| | | | 0,008 | 0,367 | 0,384 | 2,690 | 0,022 | 0,000 | 0,000 | | | | | |
| | Classical | | 0,008 | 0,367 | 0,384 | 2,690 | 0,022 | 0,000 | 0,000 | | | | | |
| | | Bottom | -0,004 | 0,283 | 0,309 | 3,510 | -0,013 | 0,000 | 0,000 | | | | | |
| | | | -0,007 | 0,268 | 0,293 | 3,510 | -0,024 | -0,001 | 0,000 | | | | | |
| | | | -0,015 | 0,328 | 0,342 | 2,200 | -0,047 | -0,002 | 0,000 | | | | | |
| | | Top | 0,076 | 1,794 | 0,840 | 2,688 | 0,042 | 0,004 | 0,000 | 8 | 15 | [5, 5] | | |
| | | | 0,074 | 1,879 | 0,840 | 2,688 | 0,039 | 0,003 | 0,000 | 8 | 15 | [10, 5] | | |
| | LSTM | | 0,071 | 1,958 | 0,857 | 2,690 | 0,036 | 0,003 | 0,000 | 16 | 15 | [5] | | |
| 90 | | Bottom | 0,052 | 1,851 | 0,839 | 2,690 | 0,028 | 0,002 | 0,000 | 8 | 15 | [5, 10] | | |
| | | | 0,050 | 1,928 | 0,840 | 2,690 | 0,026 | 0,002 | 0,000 | 8 | 20 | [10] | | |
| | | | 0,049 | 1,922 | 0,837 | 2,688 | 0,026 | 0,002 | 0,000 | 8 | 20 | [10] | | |
| | | Top | 0,055 | 1,836 | 0,859 | 2,882 | 0,030 | 0,002 | 0,000 | 16 | | 5 | True | True | False |
| | | | 0,051 | 1,915 | 0,863 | 2,934 | 0,027 | 0,002 | 0,000 | 16 | | 20 | True | False | True |
| | GPVAR | | 0,049 | 1,903 | 0,858 | 2,833 | 0,026 | 0,001 | 0,000 | 16 | | 5 | False | False | False |
| | | Bottom | 0,043 | 1,847 | 0,861 | 2,934 | 0,023 | 0,001 | 0,000 | 16 | | 20 | True | True | False |
| | | | 0,043 | 1,847 | 0,860 | 2,934 | 0,023 | 0,001 | 0,000 | 16 | | 10 | False | False | False |
| | | | 0,043 | 1,843 | 0,857 | 2,699 | 0,023 | 0,001 | 0,000 | 16 | | 10 | True | True | True |
| | | Top | 0,053 | 1,880 | 0,853 | 2,690 | 0,028 | 0,002 | 0,000 | 16 | | 20 | True | True | False |
| | | | 0,052 | 1,923 | 0,854 | 2,690 | 0,027 | 0,002 | 0,000 | 16 | | 15 | False | False | False |
| | VAR | | 0,049 | 1,918 | 0,855 | 2,690 | 0,025 | 0,001 | 0,000 | 16 | | 5 | True | True | True |
| | | Bottom | 0,042 | 1,829 | 0,856 | 2,690 | 0,023 | 0,001 | 0,000 | 16 | | 10 | True | False | False |
| | | | 0,041 | 1,908 | 0,851 | 2,690 | 0,022 | 0,001 | 0,000 | 16 | | 5 | False | True | True |
| | | | 0,039 | 1,914 | 0,854 | 2,699 | 0,020 | 0,001 | 0,000 | 16 | | 15 | False | False | False |
| 120 | Classical | Top | 0,005 | 0,375 | 0,395 | 2,312 | 0,012 | 0,000 | 0,000 | | | | | |

| | | | | | | | | | | | | | |
|---|---|---|---|---|---|---|---|---|---|---|---|---|---|
| | | | 0,005 | 0,375 | 0,395 | 2,312 | 0,012 | 0,000 | 0,000 | | | | |
| | | | 0,004 | 0,324 | 0,372 | 2,312 | 0,012 | 0,000 | 0,000 | | | | |
| | | Bottom | -0,002 | 0,304 | 0,365 | 2,438 | -0,006 | 0,000 | 0,000 | | | | |
| | | | -0,007 | 0,280 | 0,323 | 3,112 | -0,025 | -0,001 | 0,000 | | | | |
| | | | -0,025 | 0,343 | 0,418 | 1,868 | -0,072 | -0,004 | 0,000 | | | | |
| LSTM | Top | | 0,046 | 1,905 | 0,885 | 3,501 | 0,024 | 0,001 | 0,000 | 16 | 20 | [5] | |
| | | | 0,045 | 1,758 | 0,863 | 3,501 | 0,026 | 0,001 | 0,000 | 16 | 20 | [15] | |
| | | | 0,044 | 1,959 | 0,872 | 2,208 | 0,023 | 0,001 | 0,000 | 16 | 20 | [10] | |
| | Bottom | | 0,018 | 2,201 | 0,951 | 2,192 | 0,008 | 0,000 | 0,000 | 8 | 15 | [5, 10] | |
| | | | 0,018 | 2,123 | 0,951 | 2,192 | 0,008 | 0,000 | 0,000 | 16 | 15 | [20] | |
| | | | 0,016 | 2,140 | 0,951 | 2,192 | 0,008 | 0,000 | 0,000 | 8 | 20 | [10] | |
| GPVAR | Top | | 0,035 | 1,911 | 0,945 | 2,874 | 0,018 | 0,001 | 0,000 | 16 | | 20 | True | False | True |
| | | | 0,035 | 1,916 | 0,945 | 2,825 | 0,018 | 0,001 | 0,000 | 16 | | 20 | False | False | True |
| | | | 0,033 | 1,921 | 0,944 | 2,926 | 0,017 | 0,001 | 0,000 | 16 | | 5 | False | False | True |
| | Bottom | | 0,027 | 1,996 | 0,955 | 2,789 | 0,013 | 0,000 | 0,000 | 16 | | 10 | False | True | False |
| | | | 0,027 | 1,974 | 0,959 | 2,778 | 0,013 | 0,000 | 0,000 | 16 | | 20 | False | False | False |
| | | | 0,025 | 2,021 | 0,958 | 2,786 | 0,012 | 0,000 | 0,000 | 16 | | 20 | True | False | False |
| VAR | Top | | 0,031 | 2,003 | 0,943 | 2,833 | 0,016 | 0,001 | 0,000 | 16 | | 10 | True | True | False |
| | | | 0,031 | 2,009 | 0,944 | 2,833 | 0,015 | 0,001 | 0,000 | 16 | | 20 | False | True | False |
| | | | 0,031 | 2,019 | 0,944 | 2,833 | 0,015 | 0,001 | 0,000 | 16 | | 5 | True | True | False |
| | Bottom | | 0,026 | 2,098 | 0,957 | 2,786 | 0,013 | 0,000 | 0,000 | 16 | | 15 | True | True | True |
| | | | 0,025 | 2,088 | 0,960 | 2,786 | 0,012 | 0,000 | 0,000 | 16 | | 20 | True | True | False |
| | | | 0,024 | 2,109 | 0,961 | 2,786 | 0,011 | 0,000 | 0,000 | 16 | | 15 | True | False | False |

*Note:* Parameters specifications that did not converge were removed, since such strategies could not be backtested.

Table A.6. Strategies performance metrics and parameters for 3 best and 3 worst strategies based on a 60-days window.

| Rebalancing | Covariance Model | Strategy Rank | aRC | aSD | MD | MLD | IR | IR2 | IR3 | Batch Size | Length | Cells | Scaling | Copula | Low Rank |
|---|---|---|---|---|---|---|---|---|---|---|---|---|---|---|---|
| 30 | Classical | Top | -0,011 | 0,363 | 0,351 | 1,718 | -0,031 | -0,001 | 0,000 | | | | | | |
| | | | -0,011 | 0,363 | 0,351 | 1,718 | -0,031 | -0,001 | 0,000 | | | | | | |
| | | | -0,018 | 0,345 | 0,348 | 1,721 | -0,051 | -0,003 | 0,000 | | | | | | |
| | | Bottom | -0,053 | 0,375 | 0,405 | 1,723 | -0,141 | -0,018 | -0,001 | | | | | | |
| | | | -0,054 | 0,345 | 0,348 | 1,721 | -0,157 | -0,024 | -0,001 | | | | | | |
| | | | -0,088 | 0,358 | 0,382 | 1,726 | -0,245 | -0,057 | -0,003 | | | | | | |
| | LSTM | Top | 0,026 | 3,048 | 0,808 | 0,808 | 0,008 | 0,000 | 0,000 | 16 | 15 | [20, 20] | | | |
| | | | 0,024 | 2,804 | 0,895 | 1,400 | 0,009 | 0,000 | 0,000 | 8 | 15 | [15] | | | |
| | | | 0,024 | 2,986 | 0,804 | 0,808 | 0,008 | 0,000 | 0,000 | 16 | 20 | [20, 20] | | | |
| | | Bottom | 0,005 | 2,828 | 0,898 | 0,800 | 0,002 | 0,000 | 0,000 | 8 | 20 | [5, 10] | | | |
| | | | 0,000 | 2,898 | 0,900 | 1,493 | 0,000 | 0,000 | 0,000 | 8 | 15 | [15, 15] | | | |
| | | | -0,017 | 3,202 | 0,899 | 0,800 | -0,005 | 0,000 | 0,000 | 8 | 20 | [20, 20] | | | |
| | GPVAR | Top | -0,002 | 3,901 | 0,971 | 1,874 | 0,000 | 0,000 | 0,000 | 16 | | 20 | False | False | True |
| | | | -0,070 | 3,932 | 0,973 | 1,874 | -0,018 | -0,001 | 0,000 | 16 | | 5 | True | False | True |
| | | | -0,088 | 3,814 | 0,972 | 1,874 | -0,023 | -0,002 | 0,000 | 16 | | 15 | False | True | True |
| | | Bottom | -0,133 | 3,870 | 0,972 | 1,874 | -0,034 | -0,005 | 0,000 | 16 | | 15 | False | False | True |
| | | | -0,139 | 3,546 | 0,967 | 1,874 | -0,039 | -0,006 | 0,000 | 16 | | 20 | True | False | True |
| | | | -0,139 | 3,931 | 0,973 | 1,874 | -0,035 | -0,005 | 0,000 | 16 | | 20 | True | True | False |
| | VAR | Top | -0,032 | 3,366 | 0,965 | 2,082 | -0,010 | 0,000 | 0,000 | 16 | | 15 | True | True | False |
| | | | -0,038 | 3,360 | 0,961 | 1,871 | -0,011 | 0,000 | 0,000 | 16 | | 20 | True | False | False |
| | | | -0,040 | 3,485 | 0,970 | 1,871 | -0,011 | 0,000 | 0,000 | 16 | | 10 | True | False | False |
| | | Bottom | -0,059 | 3,620 | 0,963 | 1,874 | -0,016 | -0,001 | 0,000 | 16 | | 15 | True | False | True |
| | | | -0,070 | 3,488 | 0,951 | 1,874 | -0,020 | -0,001 | 0,000 | 16 | | 10 | True | True | True |
| | | | -0,073 | 4,097 | 0,971 | 1,874 | -0,018 | -0,001 | 0,000 | 16 | | 20 | True | False | True |
| 60 | Classical | Top | -0,001 | 0,370 | 0,341 | 1,940 | -0,002 | 0,000 | 0,000 | | | | | | |
| | | | -0,001 | 0,370 | 0,341 | 1,940 | -0,002 | 0,000 | 0,000 | | | | | | |
| | | | -0,005 | 0,307 | 0,338 | 1,970 | -0,015 | 0,000 | 0,000 | | | | | | |
| | | Bottom | -0,025 | 0,298 | 0,331 | 3,178 | -0,085 | -0,006 | 0,000 | | | | | | |
| | | | -0,027 | 0,285 | 0,332 | 3,178 | -0,094 | -0,008 | 0,000 | | | | | | |
| | | | -0,046 | 0,307 | 0,377 | 2,356 | -0,150 | -0,018 | 0,000 | | | | | | |
| | LSTM | Top | 0,080 | 1,866 | 0,778 | 1,712 | 0,043 | 0,004 | 0,000 | 16 | 20 | [5, 5] | | | |
| | | | 0,075 | 1,756 | 0,778 | 1,712 | 0,043 | 0,004 | 0,000 | 16 | 20 | [10, 5] | | | |
| | | | 0,075 | 1,974 | 0,778 | 1,162 | 0,038 | 0,004 | 0,000 | 16 | 20 | [5] | | | |
| | | Bottom | 0,044 | 2,085 | 0,891 | 0,921 | 0,021 | 0,001 | 0,000 | 16 | 20 | [20, 20] | | | |
| | | | 0,039 | 2,088 | 0,895 | 0,921 | 0,019 | 0,001 | 0,000 | 16 | 15 | [20] | | | |
| | | | 0,039 | 2,092 | 0,898 | 0,901 | 0,019 | 0,001 | 0,000 | 16 | 15 | [20, 20] | | | |
| | GPVAR | Top | 0,045 | 1,951 | 0,921 | 0,893 | 0,023 | 0,001 | 0,000 | 16 | | 5 | True | False | True |
| | | | 0,043 | 1,972 | 0,921 | 0,893 | 0,022 | 0,001 | 0,000 | 16 | | 5 | True | False | False |
| | | | 0,042 | 1,954 | 0,922 | 0,893 | 0,022 | 0,001 | 0,000 | 16 | | 20 | False | True | False |
| | | Bottom | 0,037 | 1,952 | 0,921 | 0,893 | 0,019 | 0,001 | 0,000 | 16 | | 20 | False | True | True |
| | | | 0,036 | 1,938 | 0,921 | 0,893 | 0,019 | 0,001 | 0,000 | 16 | | 15 | False | False | False |
| | | | 0,036 | 1,950 | 0,921 | 0,893 | 0,018 | 0,001 | 0,000 | 16 | | 20 | True | False | False |
| | VAR | Top | 0,045 | 1,983 | 0,919 | 0,893 | 0,023 | 0,001 | 0,000 | 16 | | 5 | False | False | False |
| | | | 0,044 | 1,979 | 0,919 | 0,893 | 0,022 | 0,001 | 0,000 | 16 | | 20 | True | True | True |
| | | | 0,041 | 2,007 | 0,918 | 0,893 | 0,021 | 0,001 | 0,000 | 16 | | 10 | False | True | False |
| | | Bottom | 0,035 | 1,964 | 0,918 | 0,893 | 0,018 | 0,001 | 0,000 | 16 | | 15 | False | True | True |
| | | | 0,035 | 1,954 | 0,917 | 0,893 | 0,018 | 0,001 | 0,000 | 16 | | 15 | False | True | True |
| | | | 0,034 | 1,947 | 0,916 | 0,893 | 0,018 | 0,001 | 0,000 | 16 | | 20 | False | False | True |
| 90 | Classical | Top | 0,005 | 0,294 | 0,339 | 1,216 | 0,016 | 0,000 | 0,000 | | | | | | |
| | | | 0,004 | 0,377 | 0,387 | 1,216 | 0,012 | 0,000 | 0,000 | | | | | | |
| | | | 0,004 | 0,377 | 0,387 | 1,216 | 0,012 | 0,000 | 0,000 | | | | | | |
| | | Bottom | -0,006 | 0,278 | 0,315 | 1,970 | -0,023 | 0,000 | 0,000 | | | | | | |
| | | | -0,007 | 0,271 | 0,308 | 1,970 | -0,027 | -0,001 | 0,000 | | | | | | |
| | | | -0,007 | 0,278 | 0,311 | 1,230 | -0,027 | -0,001 | 0,000 | | | | | | |
| | LSTM | Top | -0,069 | 2,056 | 0,892 | 1,493 | -0,034 | -0,003 | 0,000 | 8 | 15 | [15] | | | |
| | | | -0,072 | 1,932 | 0,874 | 1,490 | -0,037 | -0,003 | 0,000 | 8 | 15 | [10] | | | |
| | | | -0,073 | 2,087 | 0,922 | 1,940 | -0,035 | -0,003 | 0,000 | 8 | 15 | [5, 5] | | | |
| | | Bottom | -0,112 | 2,168 | 0,938 | 1,940 | -0,052 | -0,006 | 0,000 | 16 | 15 | [20] | | | |
| | | | -0,113 | 2,181 | 0,939 | 1,942 | -0,052 | -0,006 | 0,000 | 16 | 20 | [20] | | | |
| | | | -0,122 | 2,552 | 0,971 | 1,942 | -0,048 | -0,006 | 0,000 | 8 | 15 | [20] | | | |
| | GPVAR | Top | -0,106 | 2,346 | 0,976 | 1,493 | -0,045 | -0,005 | 0,000 | 16 | | 10 | False | True | False |
| | | | -0,107 | 2,355 | 0,977 | 1,493 | -0,045 | -0,005 | 0,000 | 16 | | 15 | True | False | True |
| | | | -0,107 | 2,451 | 0,975 | 1,499 | -0,044 | -0,005 | 0,000 | 16 | | 15 | True | False | False |

| Rebalancing | Covariance Model | Strategy Rank | | aRC | aSD | MD | MLD | IR | IR2 | IR3 | Batch Size | Length | Cells | Scaling | Copula | Low Rank |
|---|---|---|---|---|---|---|---|---|---|---|---|---|---|---|---|---|
| 120 | VAR | Bottom | | -0,117 | 2,609 | 0,984 | 1,493 | -0,045 | -0,005 | 0,000 | 16 | | 10 | True | False | True |
| | | | | -0,117 | 2,627 | 0,985 | 1,493 | -0,045 | -0,005 | 0,000 | 16 | | 20 | True | True | True |
| | | | | -0,118 | 2,546 | 0,986 | 1,493 | -0,046 | -0,006 | 0,000 | 16 | | 15 | True | True | True |
| | | Top | | -0,086 | 2,204 | 0,973 | 1,493 | -0,039 | -0,003 | 0,000 | 16 | | 10 | True | True | True |
| | | | | -0,087 | 2,273 | 0,979 | 1,493 | -0,038 | -0,003 | 0,000 | 16 | | 20 | False | False | True |
| | | | | -0,088 | 2,245 | 0,975 | 1,493 | -0,039 | -0,003 | 0,000 | 16 | | 20 | True | True | True |
| | | Bottom | | -0,112 | 2,310 | 0,981 | 1,490 | -0,049 | -0,006 | 0,000 | 16 | | 5 | False | True | True |
| | | | | -0,113 | 2,284 | 0,980 | 1,490 | -0,049 | -0,006 | 0,000 | 16 | | 15 | False | True | False |
| | | | | -0,113 | 2,462 | 0,983 | 1,493 | -0,046 | -0,005 | 0,000 | 16 | | 15 | True | False | True |
| | Classical | Top | | 0,012 | 0,300 | 0,335 | 1,189 | 0,041 | 0,002 | 0,000 | | | | | | |
| | | | | 0,011 | 0,376 | 0,402 | 1,636 | 0,029 | 0,001 | 0,000 | | | | | | |
| | | | | 0,011 | 0,376 | 0,402 | 1,636 | 0,029 | 0,001 | 0,000 | | | | | | |
| | | Bottom | | -0,003 | 0,291 | 0,370 | 2,356 | -0,010 | 0,000 | 0,000 | | | | | | |
| | | | | -0,004 | 0,281 | 0,358 | 2,356 | -0,015 | 0,000 | 0,000 | | | | | | |
| | | | | -0,018 | 0,332 | 0,405 | 1,882 | -0,055 | -0,003 | 0,000 | | | | | | |
| | LSTM | Top | | 0,091 | 1,736 | 0,888 | 1,085 | 0,053 | 0,005 | 0,000 | 8 | 15 | [15] | | | |
| | | | | 0,090 | 1,826 | 0,888 | 1,532 | 0,049 | 0,005 | 0,000 | 8 | 15 | [5] | | | |
| | | | | 0,090 | 1,831 | 0,888 | 1,532 | 0,049 | 0,005 | 0,000 | 8 | 15 | [5, 5] | | | |
| | | Bottom | | 0,062 | 1,931 | 0,945 | 1,085 | 0,032 | 0,002 | 0,000 | 8 | 15 | [20, 20] | | | |
| | | | | 0,062 | 1,974 | 0,948 | 1,038 | 0,031 | 0,002 | 0,000 | 8 | 20 | [10] | | | |
| | | | | 0,062 | 1,948 | 0,946 | 0,641 | 0,032 | 0,002 | 0,000 | 8 | 20 | [20] | | | |
| | GPVAR | Top | | 0,065 | 1,886 | 0,947 | 1,099 | 0,034 | 0,002 | 0,000 | 16 | | 10 | True | True | True |
| | | | | 0,063 | 1,824 | 0,946 | 1,099 | 0,035 | 0,002 | 0,000 | 16 | | 15 | False | False | True |
| | | | | 0,063 | 1,851 | 0,947 | 1,099 | 0,034 | 0,002 | 0,000 | 16 | | 15 | False | False | False |
| | | Bottom | | 0,057 | 1,896 | 0,950 | 1,099 | 0,030 | 0,002 | 0,000 | 16 | | 20 | False | False | False |
| | | | | 0,057 | 1,891 | 0,949 | 1,099 | 0,030 | 0,002 | 0,000 | 16 | | 5 | False | True | False |
| | | | | 0,057 | 1,899 | 0,949 | 1,099 | 0,030 | 0,002 | 0,000 | 16 | | 5 | True | False | False |
| | VAR | Top | | 0,064 | 1,919 | 0,952 | 1,099 | 0,033 | 0,002 | 0,000 | 16 | | 5 | True | False | False |
| | | | | 0,064 | 1,913 | 0,950 | 1,099 | 0,033 | 0,002 | 0,000 | 16 | | 20 | False | False | True |
| | | | | 0,064 | 1,913 | 0,950 | 1,099 | 0,033 | 0,002 | 0,000 | 16 | | 15 | False | False | False |
| | | Bottom | | 0,055 | 1,929 | 0,954 | 1,099 | 0,029 | 0,002 | 0,000 | 16 | | 10 | True | False | True |
| | | | | 0,055 | 1,902 | 0,949 | 1,099 | 0,029 | 0,002 | 0,000 | 16 | | 20 | False | True | True |
| | | | | 0,055 | 1,915 | 0,951 | 1,099 | 0,029 | 0,002 | 0,000 | 16 | | 10 | False | False | True |

*Note:* Parameters specifications that did not converge were removed, since such strategies could not be backtested.

Table A.7. Strategies performance metrics and parameters for 3 best and 3 worst strategies based on a 90-days window.

| Rebalancing | Covariance Model | Strategy Rank | | aRC | aSD | MD | MLD | IR | IR2 | IR3 | Batch Size | Length | Cells | Scaling | Copula | Low Rank |
|---|---|---|---|---|---|---|---|---|---|---|---|---|---|---|---|---|
| 30 | Classical | Top | | -0,007 | 0,365 | 0,321 | 1,537 | -0,019 | 0,000 | 0,000 | | | | | | |
| | | | | -0,007 | 0,365 | 0,321 | 1,537 | -0,019 | 0,000 | 0,000 | | | | | | |
| | | | | -0,010 | 0,340 | 0,313 | 1,534 | -0,028 | -0,001 | 0,000 | | | | | | |
| | | Bottom | | -0,029 | 0,335 | 0,322 | 1,534 | -0,088 | -0,008 | 0,000 | | | | | | |
| | | | | -0,047 | 0,370 | 0,394 | 1,542 | -0,128 | -0,015 | 0,000 | | | | | | |
| | | | | -0,059 | 0,368 | 0,370 | 1,542 | -0,160 | -0,026 | -0,001 | | | | | | |
| | LSTM | Top | | 0,033 | 3,113 | 0,899 | 1,400 | 0,010 | 0,000 | 0,000 | 16 | 15 | [15] | | | |
| | | | | 0,030 | 2,626 | 0,801 | 1,400 | 0,011 | 0,000 | 0,000 | 16 | 20 | [5] | | | |
| | | | | 0,029 | 3,594 | 0,897 | 0,800 | 0,008 | 0,000 | 0,000 | 8 | 20 | [15] | | | |
| | | Bottom | | 0,016 | 17,062 | 0,898 | 0,800 | 0,001 | 0,000 | 0,000 | 16 | 20 | [20] | | | |
| | | | | 0,012 | 2,809 | 0,898 | 1,301 | 0,004 | 0,000 | 0,000 | 8 | 15 | [10] | | | |
| | | | | 0,009 | 2,873 | 0,899 | 1,310 | 0,003 | 0,000 | 0,000 | 8 | 15 | [10] | | | |
| | GPVAR | Top | | -0,003 | 3,758 | 0,952 | 1,992 | -0,001 | 0,000 | 0,000 | 16 | | 5 | False | False | False |
| | | | | -0,006 | 4,081 | 0,955 | 1,992 | -0,001 | 0,000 | 0,000 | 16 | | 15 | False | True | False |
| | | | | -0,010 | 4,371 | 0,954 | 1,992 | -0,002 | 0,000 | 0,000 | 16 | | 20 | True | False | False |
| | | Bottom | | -0,024 | 5,905 | 0,956 | 1,992 | -0,004 | 0,000 | 0,000 | 16 | | 10 | False | True | True |
| | | | | -0,025 | 30,773 | 0,958 | 1,992 | -0,001 | 0,000 | 0,000 | 16 | | 5 | True | False | False |
| | | | | -0,059 | 15,816 | 0,958 | 1,992 | -0,004 | 0,000 | 0,000 | 16 | | 5 | False | False | True |
| | VAR | Top | | 0,001 | 3,341 | 0,951 | 1,992 | 0,000 | 0,000 | 0,000 | 16 | | 15 | False | True | False |
| | | | | -0,002 | 4,587 | 0,947 | 1,992 | 0,000 | 0,000 | 0,000 | 16 | | 15 | True | False | False |
| | | | | -0,012 | 3,701 | 0,944 | 1,992 | -0,003 | 0,000 | 0,000 | 16 | | 5 | False | False | False |
| | | Bottom | | -0,034 | 7,141 | 0,950 | 1,992 | -0,005 | 0,000 | 0,000 | 16 | | 5 | True | True | True |
| | | | | -0,035 | 34,155 | 0,959 | 1,992 | -0,001 | 0,000 | 0,000 | 16 | | 20 | False | True | True |

| | | | | | | | | | | | | | | |
|---|---|---|---|---|---|---|---|---|---|---|---|---|---|---|
| | | | -0,036 | 26,234 | 0,958 | 1,992 | -0,001 | 0,000 | 0,000 | 16 | | 10 | False | True | True |
| 60 | Classical | Top | 0,000 | 0,355 | 0,319 | 1,455 | 0,001 | 0,000 | 0,000 | | | | | | |
| | | | 0,000 | 0,355 | 0,319 | 1,455 | 0,001 | 0,000 | 0,000 | | | | | | |
| | | | 0,000 | 0,338 | 0,304 | 1,515 | -0,001 | 0,000 | 0,000 | | | | | | |
| | | Bottom | -0,005 | 0,322 | 0,302 | 1,523 | -0,016 | 0,000 | 0,000 | | | | | | |
| | | | -0,007 | 0,329 | 0,312 | 1,523 | -0,022 | 0,000 | 0,000 | | | | | | |
| | | | -0,030 | 0,346 | 0,358 | 1,504 | -0,088 | -0,007 | 0,000 | | | | | | |
| | LSTM | Top | 0,024 | 2,746 | 0,859 | 1,296 | 0,009 | 0,000 | 0,000 | 8 | 20 | [15] | | | |
| | | | 0,024 | 2,066 | 0,878 | 1,989 | 0,011 | 0,000 | 0,000 | 16 | 15 | [15] | | | |
| | | | 0,023 | 2,328 | 0,851 | 2,811 | 0,010 | 0,000 | 0,000 | 16 | 15 | [10, 5] | | | |
| | | Bottom | -0,004 | 3,551 | 0,920 | 1,510 | -0,001 | 0,000 | 0,000 | 16 | 20 | [20] | | | |
| | | | -0,004 | 3,452 | 0,920 | 1,992 | -0,001 | 0,000 | 0,000 | 16 | 15 | [20] | | | |
| | | | -0,004 | 3,593 | 0,920 | 1,992 | -0,001 | 0,000 | 0,000 | 16 | 15 | [20, 20] | | | |
| | GPVAR | Top | -0,006 | 4,285 | 0,942 | 1,992 | -0,001 | 0,000 | 0,000 | 16 | | 15 | True | False | True |
| | | | -0,013 | 5,015 | 0,944 | 1,992 | -0,003 | 0,000 | 0,000 | 16 | | 20 | True | False | True |
| | | | -0,013 | 5,021 | 0,944 | 1,992 | -0,003 | 0,000 | 0,000 | 16 | | 20 | False | True | False |
| | | Bottom | -0,014 | 5,153 | 0,944 | 1,992 | -0,003 | 0,000 | 0,000 | 16 | | 5 | False | False | False |
| | | | -0,014 | 5,113 | 0,944 | 1,992 | -0,003 | 0,000 | 0,000 | 16 | | 20 | True | True | True |
| | | | -0,015 | 5,036 | 0,944 | 1,992 | -0,003 | 0,000 | 0,000 | 16 | | 15 | True | True | True |
| | VAR | Top | -0,002 | 3,852 | 0,940 | 1,992 | -0,001 | 0,000 | 0,000 | 16 | | 5 | False | False | True |
| | | | -0,003 | 3,500 | 0,934 | 1,989 | -0,001 | 0,000 | 0,000 | 16 | | 20 | False | True | False |
| | | | -0,004 | 3,929 | 0,939 | 1,989 | -0,001 | 0,000 | 0,000 | 16 | | 15 | True | True | False |
| | | Bottom | -0,016 | 5,224 | 0,947 | 1,992 | -0,003 | 0,000 | 0,000 | 16 | | 20 | False | True | True |
| | | | -0,016 | 5,316 | 0,950 | 1,992 | -0,003 | 0,000 | 0,000 | 16 | | 10 | False | False | True |
| | | | -0,016 | 5,353 | 0,954 | 1,992 | -0,003 | 0,000 | 0,000 | 16 | | 10 | True | True | True |
| 90 | Classical | Top | 0,002 | 0,362 | 0,356 | 1,392 | 0,004 | 0,000 | 0,000 | | | | | | |
| | | | 0,002 | 0,362 | 0,356 | 1,392 | 0,004 | 0,000 | 0,000 | | | | | | |
| | | | 0,001 | 0,351 | 0,348 | 1,392 | 0,002 | 0,000 | 0,000 | | | | | | |
| | | Bottom | -0,006 | 0,345 | 0,335 | 1,392 | -0,017 | 0,000 | 0,000 | | | | | | |
| | | | -0,012 | 0,351 | 0,321 | 1,392 | -0,036 | -0,001 | 0,000 | | | | | | |
| | | | -0,021 | 0,397 | 0,388 | 1,392 | -0,053 | -0,003 | 0,000 | | | | | | |
| | LSTM | Top | 0,048 | 2,306 | 0,825 | 1,041 | 0,021 | 0,001 | 0,000 | 8 | 20 | [5] | | | |
| | | | 0,048 | 2,307 | 0,825 | 1,041 | 0,021 | 0,001 | 0,000 | 8 | 20 | [10, 5] | | | |
| | | | 0,048 | 2,307 | 0,825 | 1,041 | 0,021 | 0,001 | 0,000 | 8 | 20 | [5, 5] | | | |
| | | Bottom | 0,010 | 2,204 | 0,951 | 1,041 | 0,005 | 0,000 | 0,000 | 8 | 15 | [20, 20] | | | |
| | | | 0,010 | 2,237 | 0,953 | 1,041 | 0,004 | 0,000 | 0,000 | 8 | 20 | [20] | | | |
| | | | 0,009 | 2,251 | 0,953 | 1,025 | 0,004 | 0,000 | 0,000 | 8 | 20 | [20, 20] | | | |
| | GPVAR | Top | 0,017 | 2,070 | 0,948 | 1,047 | 0,008 | 0,000 | 0,000 | 16 | | 10 | False | True | False |
| | | | 0,017 | 2,062 | 0,947 | 1,047 | 0,008 | 0,000 | 0,000 | 16 | | 15 | False | False | False |
| | | | 0,017 | 2,065 | 0,947 | 1,047 | 0,008 | 0,000 | 0,000 | 16 | | 20 | True | False | True |
| | | Bottom | 0,008 | 1,981 | 0,949 | 1,047 | 0,004 | 0,000 | 0,000 | 16 | | 5 | True | False | False |
| | | | 0,008 | 1,983 | 0,949 | 1,047 | 0,004 | 0,000 | 0,000 | 16 | | 5 | True | False | True |
| | | | 0,008 | 2,004 | 0,949 | 1,047 | 0,004 | 0,000 | 0,000 | 16 | | 5 | False | False | False |
| | VAR | Top | 0,009 | 1,989 | 0,950 | 1,047 | 0,004 | 0,000 | 0,000 | 16 | | 10 | True | True | True |
| | | | 0,009 | 2,001 | 0,950 | 1,047 | 0,004 | 0,000 | 0,000 | 16 | | 10 | False | False | True |
| | | | 0,008 | 1,958 | 0,950 | 1,047 | 0,004 | 0,000 | 0,000 | 16 | | 20 | True | False | False |
| | | Bottom | 0,007 | 1,998 | 0,948 | 1,047 | 0,003 | 0,000 | 0,000 | 16 | | 10 | True | False | True |
| | | | 0,007 | 1,946 | 0,948 | 1,047 | 0,003 | 0,000 | 0,000 | 16 | | 20 | True | True | True |
| | | | 0,006 | 2,000 | 0,948 | 1,047 | 0,003 | 0,000 | 0,000 | 16 | | 15 | False | True | True |
| 120 | Classical | Top | 0,003 | 0,358 | 0,359 | 1,310 | 0,008 | 0,000 | 0,000 | | | | | | |
| | | | 0,003 | 0,358 | 0,359 | 1,310 | 0,008 | 0,000 | 0,000 | | | | | | |
| | | | -0,001 | 0,391 | 0,421 | 1,304 | -0,003 | 0,000 | 0,000 | | | | | | |
| | | Bottom | -0,019 | 0,320 | 0,351 | 1,299 | -0,060 | -0,003 | 0,000 | | | | | | |
| | | | -0,020 | 0,338 | 0,371 | 1,299 | -0,059 | -0,003 | 0,000 | | | | | | |
| | | | -0,037 | 0,331 | 0,383 | 1,304 | -0,113 | -0,011 | 0,000 | | | | | | |
| | LSTM | Top | 0,107 | 2,098 | 0,919 | 1,992 | 0,051 | 0,006 | 0,000 | 8 | 15 | [20] | | | |
| | | | 0,106 | 1,977 | 0,915 | 1,101 | 0,054 | 0,006 | 0,001 | 8 | 20 | [15] | | | |
| | | | 0,105 | 1,984 | 0,915 | 1,101 | 0,053 | 0,006 | 0,001 | 8 | 20 | [15, 15] | | | |
| | | Bottom | 0,060 | 2,050 | 0,932 | 1,466 | 0,029 | 0,002 | 0,000 | 16 | 20 | [20, 20] | | | |
| | | | 0,047 | 1,844 | 0,922 | 1,466 | 0,026 | 0,001 | 0,000 | 16 | 15 | [10] | | | |
| | | | 0,047 | 1,845 | 0,922 | 1,466 | 0,026 | 0,001 | 0,000 | 16 | 15 | [5, 10] | | | |
| | GPVAR | Top | 0,069 | 1,928 | 0,943 | 1,466 | 0,036 | 0,003 | 0,000 | 16 | | 10 | True | True | True |
| | | | 0,068 | 1,935 | 0,943 | 1,466 | 0,035 | 0,003 | 0,000 | 16 | | 10 | False | False | True |
| | | | 0,068 | 1,920 | 0,943 | 1,466 | 0,035 | 0,003 | 0,000 | 16 | | 15 | True | False | False |
| | | Bottom | 0,061 | 1,931 | 0,943 | 1,466 | 0,031 | 0,002 | 0,000 | 16 | | 20 | False | False | True |
| | | | 0,059 | 1,909 | 0,943 | 1,466 | 0,031 | 0,002 | 0,000 | 16 | | 20 | False | True | True |
| | | | 0,055 | 2,104 | 0,952 | 1,466 | 0,026 | 0,002 | 0,000 | 16 | | 5 | False | True | False |
| | VAR | Top | 0,079 | 1,883 | 0,943 | 1,466 | 0,042 | 0,004 | 0,000 | 16 | | 5 | False | True | True |
| | | | 0,073 | 1,825 | 0,942 | 0,737 | 0,040 | 0,003 | 0,000 | 16 | | 15 | False | False | False |
| | | | 0,069 | 1,916 | 0,943 | 1,466 | 0,036 | 0,003 | 0,000 | 16 | | 15 | False | False | True |
| | | Bottom | 0,063 | 2,099 | 0,952 | 1,466 | 0,030 | 0,002 | 0,000 | 16 | | 15 | True | False | True |
| | | | 0,060 | 2,099 | 0,952 | 1,466 | 0,029 | 0,002 | 0,000 | 16 | | 20 | False | False | True |

| | | | 0,058 | 2,101 | 0,952 | 1,466 | 0,028 | 0,002 | 0,000 | 16 | | 20 | False | True | True |

*Note:* Parameters specifications that did not converge were removed, since such strategies could not be backtested.

Table A.8. Strategies performance metrics and parameters for 3 best and 3 worst strategies based on a 120-days window.

| Rebalancing | Covariance Model | Strategy Rank | aRC | aSD | MD | MLD | IR | IR2 | IR3 | Batch Size | Length | Cells | Scaling | Copula | Low Rank |
|---|---|---|---|---|---|---|---|---|---|---|---|---|---|---|---|
| 30 | Classical | Top | 0,002 | 0,365 | 0,322 | 1,537 | 0,005 | 0,000 | 0,000 | | | | | | |
| | | | 0,002 | 0,365 | 0,322 | 1,537 | 0,005 | 0,000 | 0,000 | | | | | | |
| | | | -0,003 | 0,328 | 0,307 | 1,534 | -0,008 | 0,000 | 0,000 | | | | | | |
| | | Bottom | -0,022 | 0,326 | 0,304 | 1,534 | -0,066 | -0,005 | 0,000 | | | | | | |
| | | | -0,043 | 0,356 | 0,387 | 1,542 | -0,121 | -0,013 | 0,000 | | | | | | |
| | | | -0,047 | 0,331 | 0,309 | 1,542 | -0,142 | -0,022 | -0,001 | | | | | | |
| | LSTM | Top | 0,046 | 2,699 | 0,804 | 0,934 | 0,017 | 0,001 | 0,000 | 8 | 20 | [5, 5] | | | |
| | | | 0,046 | 2,774 | 0,804 | 0,934 | 0,016 | 0,001 | 0,000 | 8 | 20 | [10, 5] | | | |
| | | | 0,046 | 2,775 | 0,804 | 0,934 | 0,016 | 0,001 | 0,000 | 8 | 20 | [10, 5] | | | |
| | | Bottom | 0,020 | 59,566 | 0,897 | 0,595 | 0,000 | 0,000 | 0,000 | 8 | 20 | [15] | | | |
| | | | 0,019 | 3,284 | 0,899 | 1,400 | 0,006 | 0,000 | 0,000 | 16 | 20 | [5, 5] | | | |
| | | | 0,018 | 2,833 | 0,898 | 0,507 | 0,006 | 0,000 | 0,000 | 8 | 15 | [15] | | | |
| | GPVAR | Top | 0,017 | 3,031 | 0,925 | 1,992 | 0,006 | 0,000 | 0,000 | 16 | | 5 | True | False | False |
| | | | 0,015 | 3,091 | 0,932 | 1,992 | 0,005 | 0,000 | 0,000 | 16 | | 5 | False | False | True |
| | | | 0,010 | 3,216 | 0,936 | 1,992 | 0,003 | 0,000 | 0,000 | 16 | | 15 | True | True | True |
| | | Bottom | -0,003 | 3,612 | 0,943 | 1,992 | -0,001 | 0,000 | 0,000 | 16 | | 15 | False | False | True |
| | | | -0,003 | 3,629 | 0,943 | 1,992 | -0,001 | 0,000 | 0,000 | 16 | | 15 | True | False | True |
| | | | -0,003 | 3,614 | 0,943 | 1,992 | -0,001 | 0,000 | 0,000 | 16 | | 10 | True | True | True |
| | VAR | Top | 0,011 | 3,053 | 0,929 | 1,992 | 0,003 | 0,000 | 0,000 | 16 | | 15 | True | False | True |
| | | | 0,010 | 3,016 | 0,928 | 1,992 | 0,003 | 0,000 | 0,000 | 16 | | 20 | True | True | True |
| | | | 0,009 | 3,117 | 0,930 | 1,992 | 0,003 | 0,000 | 0,000 | 16 | | 10 | True | False | True |
| | | Bottom | 0,001 | 3,809 | 0,931 | 1,992 | 0,000 | 0,000 | 0,000 | 16 | | 10 | True | False | False |
| | | | -0,001 | 4,039 | 0,935 | 1,992 | 0,000 | 0,000 | 0,000 | 16 | | 20 | False | True | True |
| | | | -0,028 | 5,182 | 0,937 | 1,992 | -0,005 | 0,000 | 0,000 | 16 | | 10 | False | False | False |
| 60 | Classical | Top | 0,010 | 0,346 | 0,325 | 1,148 | 0,028 | 0,001 | 0,000 | | | | | | |
| | | | 0,010 | 0,346 | 0,325 | 1,148 | 0,028 | 0,001 | 0,000 | | | | | | |
| | | | 0,009 | 0,277 | 0,308 | 2,356 | 0,031 | 0,001 | 0,000 | | | | | | |
| | | Bottom | -0,005 | 0,274 | 0,298 | 2,688 | -0,017 | 0,000 | 0,000 | | | | | | |
| | | | -0,008 | 0,311 | 0,318 | 1,970 | -0,025 | -0,001 | 0,000 | | | | | | |
| | | | -0,025 | 0,284 | 0,326 | 2,356 | -0,088 | -0,007 | 0,000 | | | | | | |
| | LSTM | Top | 0,094 | 2,054 | 0,860 | 1,041 | 0,046 | 0,005 | 0,000 | 16 | 20 | [5, 5] | | | |
| | | | 0,093 | 2,031 | 0,772 | 1,041 | 0,046 | 0,006 | 0,000 | 16 | 20 | [5] | | | |
| | | | 0,093 | 1,926 | 0,860 | 1,041 | 0,048 | 0,005 | 0,000 | 16 | 20 | [10, 10] | | | |
| | | Bottom | 0,060 | 2,041 | 0,874 | 0,773 | 0,030 | 0,002 | 0,000 | 8 | 15 | [15] | | | |
| | | | 0,058 | 2,001 | 0,878 | 1,033 | 0,029 | 0,002 | 0,000 | 8 | 15 | [20, 20] | | | |
| | | | 0,054 | 2,026 | 0,876 | 0,773 | 0,026 | 0,002 | 0,000 | 8 | 15 | [20] | | | |
| | GPVAR | Top | 0,054 | 2,094 | 0,915 | 1,041 | 0,026 | 0,002 | 0,000 | 16 | | 5 | True | True | True |
| | | | 0,050 | 1,877 | 0,916 | 1,041 | 0,027 | 0,001 | 0,000 | 16 | | 10 | True | False | True |
| | | | 0,050 | 1,871 | 0,915 | 1,041 | 0,027 | 0,001 | 0,000 | 16 | | 10 | True | True | False |
| | | Bottom | 0,046 | 1,850 | 0,915 | 1,041 | 0,025 | 0,001 | 0,000 | 16 | | 5 | True | False | True |
| | | | 0,046 | 1,975 | 0,914 | 1,041 | 0,023 | 0,001 | 0,000 | 16 | | 20 | False | True | True |
| | | | 0,046 | 1,885 | 0,914 | 1,041 | 0,024 | 0,001 | 0,000 | 16 | | 5 | False | False | True |
| | VAR | Top | 0,063 | 1,770 | 0,807 | 1,162 | 0,036 | 0,003 | 0,000 | 16 | | 20 | False | False | True |
| | | | 0,054 | 1,889 | 0,915 | 1,041 | 0,029 | 0,002 | 0,000 | 16 | | 10 | True | True | True |
| | | | 0,052 | 1,883 | 0,915 | 1,041 | 0,028 | 0,002 | 0,000 | 16 | | 20 | True | False | False |
| | | Bottom | 0,048 | 1,817 | 0,917 | 1,041 | 0,027 | 0,001 | 0,000 | 16 | | 10 | False | False | True |
| | | | 0,047 | 1,806 | 0,916 | 1,041 | 0,026 | 0,001 | 0,000 | 16 | | 15 | False | False | True |
| | | | 0,045 | 1,823 | 0,916 | 1,041 | 0,025 | 0,001 | 0,000 | 16 | | 10 | False | True | True |
| 90 | Classical | Top | 0,015 | 0,291 | 0,281 | 2,526 | 0,051 | 0,003 | 0,000 | | | | | | |
| | | | 0,013 | 0,357 | 0,384 | 1,496 | 0,036 | 0,001 | 0,000 | | | | | | |
| | | | 0,013 | 0,357 | 0,384 | 1,496 | 0,036 | 0,001 | 0,000 | | | | | | |
| | | Bottom | 0,000 | 0,278 | 0,280 | 2,315 | 0,000 | 0,000 | 0,000 | | | | | | |
| | | | -0,001 | 0,281 | 0,283 | 2,315 | -0,003 | 0,000 | 0,000 | | | | | | |
| | | | -0,009 | 0,293 | 0,307 | 1,304 | -0,030 | -0,001 | 0,000 | | | | | | |
| | LSTM | Top | 0,105 | 1,861 | 0,770 | 1,096 | 0,056 | 0,008 | 0,001 | 16 | 20 | [5] | | | |
| | | | 0,105 | 1,605 | 0,741 | 1,148 | 0,065 | 0,009 | 0,001 | 8 | 20 | [15, 15] | | | |
| | | | 0,103 | 1,653 | 0,762 | 1,096 | 0,062 | 0,008 | 0,001 | 8 | 15 | [15] | | | |

| | | | | | | | | | | | | | | |
|---|---|---|---|---|---|---|---|---|---|---|---|---|---|---|
| | | Bottom | 0,081 | 1,735 | 0,815 | 1,096 | 0,046 | 0,005 | 0,000 | 8 | 15 | [5, 10] | | | |
| | | | 0,080 | 1,715 | 0,810 | 1,096 | 0,047 | 0,005 | 0,000 | 16 | 15 | [20] | | | |
| | | | 0,080 | 1,745 | 0,815 | 1,096 | 0,046 | 0,004 | 0,000 | 8 | 15 | [10] | | | |
| | GPVAR | Top | 0,072 | 1,718 | 0,866 | 1,397 | 0,042 | 0,003 | 0,000 | 16 | | 5 | True | True | True |
| | | | 0,072 | 1,744 | 0,847 | 1,397 | 0,041 | 0,004 | 0,000 | 16 | | 5 | True | True | False |
| | | | 0,072 | 1,748 | 0,851 | 1,397 | 0,041 | 0,003 | 0,000 | 16 | | 20 | False | True | False |
| | | Bottom | 0,069 | 1,809 | 0,835 | 1,395 | 0,038 | 0,003 | 0,000 | 16 | | 15 | False | False | True |
| | | | 0,067 | 1,813 | 0,835 | 1,397 | 0,037 | 0,003 | 0,000 | 16 | | 20 | False | False | True |
| | | | 0,064 | 1,627 | 0,852 | 1,397 | 0,039 | 0,003 | 0,000 | 16 | | 5 | False | True | True |
| | VAR | Top | 0,072 | 1,688 | 0,857 | 1,397 | 0,043 | 0,004 | 0,000 | 16 | | 10 | False | False | False |
| | | | 0,068 | 1,592 | 0,832 | 1,397 | 0,043 | 0,003 | 0,000 | 16 | | 15 | False | False | True |
| | | | 0,068 | 1,573 | 0,854 | 1,397 | 0,043 | 0,003 | 0,000 | 16 | | 20 | True | True | True |
| | | Bottom | 0,064 | 1,571 | 0,856 | 1,397 | 0,041 | 0,003 | 0,000 | 16 | | 15 | False | False | False |
| | | | 0,064 | 1,630 | 0,822 | 1,397 | 0,039 | 0,003 | 0,000 | 16 | | 20 | True | False | False |
| | | | 0,064 | 1,574 | 0,856 | 1,397 | 0,041 | 0,003 | 0,000 | 16 | | 10 | True | True | False |
| 120 | Classical | Top | 0,011 | 0,321 | 0,317 | 1,973 | 0,033 | 0,001 | 0,000 | | | | | | |
| | | | 0,011 | 0,321 | 0,317 | 1,973 | 0,033 | 0,001 | 0,000 | | | | | | |
| | | | 0,003 | 0,295 | 0,310 | 2,526 | 0,011 | 0,000 | 0,000 | | | | | | |
| | | Bottom | -0,007 | 0,289 | 0,302 | 2,526 | -0,024 | -0,001 | 0,000 | | | | | | |
| | | | -0,007 | 0,284 | 0,300 | 2,526 | -0,026 | -0,001 | 0,000 | | | | | | |
| | | | -0,034 | 0,307 | 0,285 | 2,526 | -0,111 | -0,013 | 0,000 | | | | | | |
| | LSTM | Top | -0,051 | 1,771 | 0,798 | 1,167 | -0,029 | -0,002 | 0,000 | 16 | 20 | [10, 10] | | | |
| | | | -0,058 | 1,603 | 0,794 | 0,852 | -0,036 | -0,003 | 0,000 | 8 | 20 | [20, 20] | | | |
| | | | -0,059 | 1,356 | 0,752 | 1,208 | -0,044 | -0,003 | 0,000 | 16 | 15 | [10] | | | |
| | | Bottom | -0,097 | 1,653 | 0,904 | 0,852 | -0,059 | -0,006 | -0,001 | 8 | 15 | [10, 10] | | | |
| | | | -0,097 | 1,653 | 0,904 | 0,852 | -0,059 | -0,006 | -0,001 | 8 | 15 | [5, 10] | | | |
| | | | -0,098 | 1,654 | 0,915 | 0,852 | -0,059 | -0,006 | -0,001 | 16 | 20 | [10] | | | |
| | GPVAR | Top | -0,091 | 1,541 | 0,910 | 0,852 | -0,059 | -0,006 | -0,001 | 16 | | 5 | False | False | True |
| | | | -0,091 | 1,572 | 0,909 | 0,852 | -0,058 | -0,006 | -0,001 | 16 | | 5 | True | False | True |
| | | | -0,093 | 1,567 | 0,910 | 0,852 | -0,059 | -0,006 | -0,001 | 16 | | 5 | True | True | True |
| | | Bottom | -0,094 | 1,579 | 0,910 | 0,852 | -0,059 | -0,006 | -0,001 | 16 | | 5 | False | False | False |
| | | | -0,095 | 1,637 | 0,910 | 0,852 | -0,058 | -0,006 | -0,001 | 16 | | 15 | True | False | True |
| | | | -0,095 | 1,655 | 0,910 | 0,852 | -0,058 | -0,006 | -0,001 | 16 | | 10 | False | False | True |
| | VAR | Top | -0,104 | 1,553 | 0,935 | 0,852 | -0,067 | -0,007 | -0,001 | 16 | | 15 | True | True | False |
| | | | -0,104 | 1,553 | 0,936 | 0,852 | -0,067 | -0,007 | -0,001 | 16 | | 5 | False | True | True |
| | | | -0,104 | 1,552 | 0,935 | 0,852 | -0,067 | -0,007 | -0,001 | 16 | | 20 | True | False | False |
| | | Bottom | -0,107 | 1,731 | 0,933 | 0,852 | -0,062 | -0,007 | -0,001 | 16 | | 20 | True | False | True |
| | | | -0,107 | 1,722 | 0,929 | 0,852 | -0,062 | -0,007 | -0,001 | 16 | | 15 | False | True | True |
| | | | -0,107 | 1,729 | 0,932 | 0,852 | -0,062 | -0,007 | -0,001 | 16 | | 20 | True | True | True |

*Note:* Parameters specifications that did not converge were removed, since such strategies could not be backtested.